\documentclass[aps, pra, a4paper, showpacs, twocolumn,10pt, superscriptaddress]{revtex4-1}
\usepackage{amssymb,bbm,bm,textcomp, nicefrac,geometry, epstopdf}
\usepackage{graphicx}
\usepackage{amsmath}
\usepackage{amsthm}
\usepackage{color}
\usepackage{bbding}
\usepackage{enumerate}
\usepackage[usenames,dvipsnames]{xcolor}
\usepackage{subfigure}
\usepackage{float}
\usepackage[mathscr]{eucal}
\usepackage{esvect}
\usepackage[toc,page]{appendix}
\def\ba{\begin{aligned}}
\def\ea{\end{aligned}}
\def\be{\begin{equation}}
\def\ee{\end{equation}}
\def\bestar{\begin{equation*}}
\def\eestar{\end{equation*}}
\def\bea{\begin{eqnarray}}
\def\eea{\end{eqnarray}}
\def\bi{\begin{itemize}}
\def\ei{\end{itemize}}
\def\bc{\begin{center}}
\def\ec{\end{center}}
\def\bt{\begin{tabular}}
\def\et{\end{tabular}}
\def\btm{\begin{theorem}}
\def\etm{\end{theorem}}
\def\barr{\begin{array}}
\def\earr{\end{array}}
\def\bbmat{\begin{bmatrix}}
\def\ebmat{\end{bmatrix}}
\def\bpmat{\begin{pmatrix}}
\def\epmat{\end{pmatrix}}
\def\Order{\mathcal O}
\def\one{\mbox{$1 \hspace{-1.0mm}  {\bf l}$}}

\newcommand{\expB}{\langle \mathcal{B} \rangle}
\newcommand{\matB}{\mathcal{B}}

\newcommand{\ketzero}{\ket{\boldsymbol{0}}}

\def\adj#1{#1^{\dagger}}
\def\ket#1{\left| #1\right>} 							% | var1 >
\def\bra#1{\left< #1\right|}							% < var1 |
\def\bk#1#2{\left< #1\middle| #2 \right>}				% < var1 | var2 >
\def\kb#1#2{\ket{#1}\bra{#2}}  							% | var1 >< var2 |
\def\bmk#1#2#3{\left< #1\middle| #2 \middle| #3 \right>}  % < var1 | var2 | var3 >
\def\proj#1{\ket{#1}\bra{#1}}							% | var1 >< var1 |
\newtheorem{theorem}{Theorem}

\newcommand{\expect}[1]{\langle #1 \rangle}

\newcommand{\trans}[1]{{#1}^{\mathsf{T}}}
\newcommand\defn[1]{\textsl{#1}}
\newcommand{\eq}[1]{Eq.~(\ref{eq.#1})}
\newcommand{\sect}[1]{Sec.~\ref{sect.#1}}
\newcommand{\figu}[1]{Fig.~\ref{fig.#1}}
\begin{document}
\date{\today}

\title{Compressed quantum metrology for the Ising Hamiltonian}

\author{W.~L.~Boyajian}
\affiliation{Institut f\"ur Theoretische Physik, Universit\"at Innsbruck, Technikerstr. 25, A-6020 Innsbruck, Austria}
\author{M.~Skotiniotis}
\affiliation{F\'isica Te\`orica: Informaci\'o i Fen\`omens Qu\`antics, Departament de F\'isica, Universitat Aut\`onoma de Barcelona, 08193 Bellatera (Barcelona) Spain}
\author{W.~D\"ur}
\affiliation{Institut f\"ur Theoretische Physik, Universit\"at Innsbruck, Technikerstr. 25, A-6020 Innsbruck, Austria}
\author{B.~Kraus}
\affiliation{Institut f\"ur Theoretische Physik, Universit\"at Innsbruck, Technikerstr. 25, A-6020 Innsbruck, Austria}

\begin{abstract}
We show how quantum metrology protocols that seek to estimate the parameters of a Hamiltonian that exhibits a quantum 
phase transition can be efficiently simulated on an exponentially smaller quantum computer.  Specifically, by exploiting the fact 
that the ground state of such a Hamiltonian changes drastically around its phase transition point, we construct a 
suitable observable from which one can estimate the relevant parameters of the Hamiltonian with Heisenberg scaling precision.
We then show how, for the one-dimensional Ising Hamiltonian with transverse magnetic field acting on $N$ spins, such a 
metrology protocol can be efficiently simulated on an exponentially smaller quantum computer while maintaining the same 
Heisenberg scaling, i.e., $\Order (N^{-2})$ precision and derive the explicit circuit that accomplishes the simulation.  
\end{abstract}
\pacs{03.67.-a, 03.65.Ud, 03.65.Yz, 03.65.Ta}
\maketitle

\section{Introduction}
Determining the properties and parameters of physical systems is a central problem in many experiments. Quantum metrology 
deals with the question of how this can be achieved optimally using a given number of resources, counted in terms of number 
of systems or evolution time~\cite{Huelga:97, GLM04,Giovanetti:11, Demkowicz:15}. It is a key result that quantum mechanics 
offers a significant advantage for achieving this task. In fact, using entangled probe states one finds that a quadratic gain in 
precession as compared to any classical strategy can be achieved~\cite{GLM04, Giovanetti:11}, provided that one deals with 
noiseless evolutions~\cite{Huelga:97, Escher:11, Kolodynski:12, Kolodynski:13, Sekatski:16}.

We consider the problem of estimating some unknown parameter of a strongly coupled system, e.g., the interaction strength of 
a 1D chain of $N$ interacting spins. The physics of such strongly coupled systems is widely studied~\cite{Sachdev:07}. 
Recently it was realized that certain classes of such strongly interacting systems can be simulated by an exponentially smaller 
system in a compressed way, where relevant properties such as quantum phase transitions or two-point correlation functions 
can be determined efficiently~\cite{Kraus:11,Boyajian:13,Boyajian:15}.

Here we show that one can also perform quantum metrology on such a compressed system. In our approach, we assume that 
we have access to a family of Hamiltonians describing $N$ interacting spins for different system sizes $N$, and can switch the 
Hamiltonian on and off at will. We illustrate our approach with the one-dimensional Ising model with transversal magnetic field, 
where the goal is to determine the unknown coupling strength of the model. We remark that the estimation of the coupling 
strength does not fall into the widely studied class of problems where the unknown parameter is a multiplicative 
constant~\cite{GLM04, Giovanetti:11, Demkowicz:15, Escher:11, Kolodynski:12, Kolodynski:13, Sekatski:16}, but is of a more 
complicated type where the overall Hamiltonian is some function of the parameter~\cite{Pang:14}. Nevertheless, this problem 
has been fully solved in a standard metrology approach~\cite{Skotiniotis:15}, where the Hamiltonian is applied to some optimal 
state which is subsequently measured, and an estimate of the parameter is determined from the measurement statistics.

We provide an alternative approach that is based on estimating a certain observable near the quantum phase transition of the 
system which is first driven to the ground state, similar as in~\cite{Zanardi:08,Invernizzi:08,Mehboudi:16}. We use the 
Hamiltonian to drive the system to its ground state via adiabatic evolution, and determine the unknown parameter from its 
ground state. We show that one can do this using an exponentially smaller system. While this does not imply a super-
Heisenberg scaling when resources are properly counted, it might nevertheless be an interesting experimental alternative.
We use that the Ising model---more precisely the measurement of certain observables in this model---can be efficiently 
simulated with an exponentially smaller number of spins using matchgate circuits, as demonstrated 
in~\cite{Kraus:11,Boyajian:13,Boyajian:15}. This is due to the fact that the system can be mapped to non-interacting fermions, 
thereby reducing the effective dimension of the Hilbert space~\cite{Knill:01,Terhal:02, Valiant:02,Valiant:08,Jozsa:08,Jozsa:09}. 
We show that there exist such an observable that, when estimated at the ground state of the model close to the quantum phase 
transition, allows one to determine the coupling strength with a precession that is super-classical.

We remark that with the assumed level of control, there are alternative methods to achieve Heisenberg scaling in precision for 
the estimation of the coupling strength with limited resources. In fact, a simple sequential scheme that operates with only two 
qubits suffices. This scheme uses standard quantum metrology techniques and imprints the parameter in question as a phase 
onto a single qubit (see Sec. II A). Nevertheless, our work shows that indirect methods concerned with the fast change of 
properties near to a quantum phase transition can be used and combined with techniques from compressed quantum 
computation~\cite{Jozsa:08,Jozsa:09,Kraus:11, Boyajian:13,Boyajian:15}. This approach is not limited to the Ising model, or the 
estimation of the coupling strength, and might in fact have an advantage over standard metrological protocols when considering 
problems such as the estimation of two-point correlation functions or other ground state properties that might not be directly 
accessible by time evolution with respect to the Hamiltonian $H$. 

The paper is organized as follows. In Sec.~\ref{sect.Preliminaries} we describe the setting and give background information 
on matchgate circuits, compressed quantum computation and its relation to the Ising Hamiltonian, as well as quantum 
metrology. In Sec.~\ref{sect. Compressed_metro} we show how compressed metrology of the Ising model can be achieved. 
We first recall how to drive the compressed system to the ground state via adiabatic evolution, and then provide a 
suitable observable that can be measured on the compressed system efficiently such that the coupling 
strength of the interaction can be determined with super-classical precision. 
Surprisingly, the magnetization as an obvious candidate fails to provide such a 
super-classical scaling. We summarize and conclude in Sec.~\ref{sect.conclusion}.

%%%%%%%%%%%%%%%%%%%%%%%%%%%%%%%%%%%%%%%%%%%%%%%%%%%%%%%%%%%%%%%
%%%%%%%%%%%%%%%%%%%%%%%%%%%%%%%%%%%%%%%%%%%%%%%%%%%%%%%%%%%%%%%
\section{Background}
	\label{sect.Preliminaries}

In this section we review the main ingredients utilized throughout this work in order to realize a compressed quantum metrology 
protocol for the one dimensional Ising Hamiltonian with transverse field.  After establishing notation in 
\sect{Preliminaries-Notation} and presenting the general setting considered here (\sect{ Setting}), we review key properties 
of matchgates and matchgate circuits in \sect{Preliminaries-Matchgates}. We then specialize to the application of matchgate 
circuits for the case of the one-dimensional Ising Hamiltonian with transverse magnetic field in \sect{Preliminaries-Hamiltonian}, 
where we briefly recall how one can explicitly construct the ground state of such a Hamiltonian using matchgate 
circuits~\cite{Kraus:11, Boyajian:13}.  Finally, we review key concepts of quantum metrology, and in particular
how the phase transition exhibited by the one-dimensional Ising Hamiltonian can be used to enhance estimation precision of its 
relevant parameters in \sect{Preliminaries-Metrology}.

\subsection{Notation}
	\label{sect.Preliminaries-Notation}
We denote by $X,Y,Z$ the Pauli matrices, and by $\one$ the identity operator. The computational basis states will be denoted 
by $\ket{k}$ for $k \in \left\{0 ,1 \right\}^{\otimes N}$. We will consider a spin chain of $N=2^m$ qubits, for some integer $m$. 
We will also denote by $\ketzero $ the state of $N$ qubits $\ket{0}^{\otimes N}$. Qubits, matrix components, and fermionic 
operators are counted starting from 0, in order to unify the indexing.  Controlled operations are denoted by 
$\Lambda_{i_0, \dots , i_k } (A_j)$, representing that $A$ is applied on qubit $j$ in the exclusive case where each of the
qubits $i_0,\dots, i_k$ are in the state $\ket{1}$.  

\subsection{General setting}
	\label{sect. Setting}
We consider the situation where one has access to a general interaction, in our case the Ising interaction 
(see Eq. (\ref{eq. Hamiltonian})), for an arbitrary number of qubits.  Our aim is to estimate some parameter which defines this 
interaction, such as the coupling constant, $J$. We assume to know and have control over the local magnetic field, $B$. Using 
that the expectation value of certain observables evaluated for the ground state change abruptly at the phase transition, 
$J\approx B$, one can infer the value of $J$ by preparing the ground state of the system for certain values of $B$ and 
measuring this observable. Note that using this idea for parameter estimation has been suggested 
in~\cite{Zanardi:06, Zanardi:07, Zanardi:08, Invernizzi:08, Mehboudi:16}. 

Here, we use a different approach. 
First, instead of cooling the system to the ground state we use the fact that it can be prepared using adiabatic evolution. 
Then, we use the fact that this evolution and the measurement of certain observables is a so-called match gate circuit, 
which has been shown to be compressible to an exponentially smaller quantum computer~\cite{Kraus:11,Boyajian:13}. 
We then show that this compressed evolution can be realized using the Ising 
interaction for the unknown parameter $J$. In fact, we show that only two qubits need to interact at each step of the 
computation. We then derive an observable which fulfills both requirements, namely that it is compressible and that it allows to 
estimate the coupling parameter with super-classical precision. 

In order to state the required resources, let us mention here that the 
adiabatic evolution is discretized into a so-called digital-adiabatic evolution, i.e., into a sequence of $L+1$ time evolutions 
governed by constant Hamiltonians. In order to be able to neglect both the probability of excitations during the adiabatic 
evolution and the error due to the discretization, it suffices to chose $L$ and the total time of the adiabatic evolution, $T$, as 
polynomials in the number of qubits, $N$.
	
The compressed circuit which simulates the digital-adiabatic evolution runs on $m+2$ qubits, where 
$m=\log (N)$, and also consists of a sequence of $L+1$ steps. At each of these steps $\Order(m^2)$ elementary (fast) control 
gates are acted on the $m$ qubits and two qubits interact via an Ising Hamiltonian 
for a short time, $\delta t$, which depends on the Trotter step. The total time the system interacts governed by the Ising 
interaction is $T\equiv (L+1)\delta t$, i.e., the duration of the adiabatic evolution which is simulated.  

We note that for the particular example investigated here, there exist easier and more economical ways of estimating the 
parameter.  For instance, setting the local magnetic field $B$ equal to zero (which we assume to be able to do here), the 
remaining Hamiltonian (acting on two qubits) $J X_1\otimes X_2$ can simply be applied onto a state 
$|\varphi\rangle|0_x\rangle$. Using additional $\pi/4$ rotations along the $y$-axis on the first qubit, one can convert the 
Hamiltonian to $J Z_1\otimes X_2$, thereby imprinting the information on $J$ directly onto the phase. The choice 
$|\varphi\rangle=|0_x\rangle = (|0\rangle + |1\rangle)/\sqrt{2}$ is optimal and leads to Heisenberg scaling in precision, with 
respect to the total evolution time $T$, i.e., $\delta J^2\geq (\nu T)^{-2}$ where $\nu$ is the number of repetitions.  Note that 
there is no dependence on the system 
size, $N$, here as the evolution applies \emph{sequentially} on the two qubits~\footnote{We note that this strategy yields the 
exact same precision as the standard parallel scheme, where $N$ suitably entangled probe systems  sense the evolution in 
parallel for a short time $\delta t$ before they are measured.  Moreover the resources used are the same $N\delta t\equiv T$.}.

However, our scheme is not restricted to the Ising model. It applies to any situation where the parameter of interest can be 
estimated using a matchgate circuit as long as the required interaction in the compressed model can be realized using the 
considered Hamiltonian.  Moreover, our scheme can also be utilized to cases where the parameter can only be infered by 
measuring for instance (staggered) correlation functions something which would be impossible using the 
scheme described above.  

\subsection{Matchgate circuits and compressed quantum simulation}
	\label{sect.Preliminaries-Matchgates}
	\newlength{\myl}
	\settowidth{\myl}{$X_j$}

Matchgates are a particular set of two qubit gates satisfying certain algebraic constraints with applications in the theory of 
perfect matchings of graphs~\cite{Valiant:02, Valiant:08}, fermionic linear optics~\cite{Knill:01, Terhal:02}, as well as 
one-dimensional spin chains~\cite{Kraus:11, Boyajian:13, Boyajian:15}.

A matchgate, $G(A,B)$, is a two qubit gate of the form
\be
G(A,B)=\bbmat A_{11}&0&0&A_{12}\\
0&B_{11}&B_{12}&0\\
0&B_{21}&B_{22}&0\\
A_{21}&0&0&A_{22}
\ebmat,
\label{eq.Matchgate}
\ee
where $A_{ij}$ are the matrix elements of $A$ (and similarly for $B$). The unitary operators $A$ and $B$ act only on states 
belonging to the even and odd parity subspace of two qubits respectively and satisfy the
condition $\mathrm{det}(A)=\mathrm{det}(B)$. In what follows we will only consider matchgates (or products of them) acting on 
nearest neighbors. Hence, we call nearest neighbor matchgates in the following simply matchgates.

Any matchgate, or product of matchgates, can be written as
\be
	U = e ^{-iH } ,
	\label{eq.U matchgate}
\ee
where $H$ is a Hermitian operator that can be written as
\be
H = i \sum_{j \neq k = 0 } ^{ 2N - 1} h _{jk} x_j x_k .
\label{eq.H quadratic}
\ee
Here the matrix $h$ is a $2N\times 2N$ real anti-symmetric matrix~\cite{Jozsa:08}, and we have introduced the set of $2 N$ 
Hermitian operators
$\left\{ x_j \right\} _{j = 0 , \ldots , 2 N -1}$ with (Jordan-Wigner represenation)
\be
\ba
&x _ { 2 j }=\bigotimes_{k=0}^{j-1} Z _k \otimes \makebox[\myl] {$X_j$} \bigotimes_{k=j+1}^N  \one_ {k}, \\
&x _ { 2 j + 1 }= \bigotimes_{k=0}^{j-1} Z _k \otimes \makebox[\myl] {$Y_j$} \bigotimes_{k=j+1}^N  \one_ {k}.
\ea
\label{eq.Majorana operators}
\ee

Note that this set of operators generate the \defn{Clifford algebra},
$\mathcal{C}_{2N}$, as they satisfy $x_j^2=1$ and the anti-commutation relations
$\left\{x_j , x_ k \right\} = 2 \delta _{jk } \; \forall j,\,k$.
The important property of a matchgate, as given in \eq{U matchgate}, is that its action on any of the generators of 
$\mathcal{C}_{2N}$ can be shown to give~\cite{Jozsa:08}
\be
\adj{ U } c_j U = \sum _{k = 0 } ^ {2N - 1} R _{jk} c_k,
\label{eq.U x U}
\ee
where the matrix $R \in \mathcal{SO} (2N)$ is given explicitly by
\be
R = e^ {4 h}.
\label{eq.R}
\ee

It has been demonstrated that any circuit of matchgates satisfying the constraints: {\it (i)} matchgates act only on
\defn{nearest-neighbours},{\it (ii)} the input state of the $N$ qubits is any \defn{computational basis state} and, {\it (iii)}
the output of the circuit is the result of measuring \defn{any single qubit in the computational basis}, can be simulated
classically efficiently~\cite{Valiant:02, Terhal:02, Jozsa:08}. Let us call in the following a circuit fulfilling constraints (i)--(iii) a 
\defn{matchgate circuit} (MGC).

The reason why MGC can be classically simulated efficiently can be easily understood as follows. First, note that any MGC with
input state $\ket{k_1,\ldots k_N}$ ($k_i\in \{0,1\}$) and $Z$-measurement on qubit $k$ can be mapped to an equivalent MGC
with input state $\ket{0,\ldots 0}$ and $Z$-measurement on the first qubit~\cite{Jozsa:09}. Consider a MGC where the initial 
state, $\ketzero$, evolves under the action of a unitary $U$ given in~\eq{U matchgate} followed by a measurement of the 
observable $Z_0 = -i x_{0} x_{1}$ on the final state. Defining the $2N\times 2N$ matrix $S$ with components
$S_{jk} = \bra{\boldsymbol{0}}\left( - i x_ j  x_k \right) \ketzero$, the outcome of the aforementioned circuit can be written as
\be
\ba
\expect{Z_0}&=\bra{ \boldsymbol{0} } \adj{U} \left( - i x_{0} x_{1}\right) U \ketzero \\
&= \left[ R S \trans{R} \right] _{0,1}.
\ea
\label{eq.Z k expectation value}
\ee
As our MGC consists of a sequence of nearest neighbour matchgates, i.e., $U=U_m\cdots U_1$, the $2N\times 2N$ matrix 
$R$ can be written as $R_m \cdots R_1$, where $R_i$ is associated to $U_i$ according to~\eq{R}. Moreover, 
$S=\one\otimes i Y$, where $\one$ is the $N$-dimensional identity. Hence, the matrices as well as their product can be 
computed efficiently.

In~\cite{Jozsa:09} it has been shown that any MGC on $N$ qubits can be compressed to a universal quantum computation
running on $\log(m)+3$ qubits. This simulation is efficient as the size of the compressed computation, i.e., the number of single
and two-qubit gates, is ${\cal O} \left[ M\log(N) \right]$, if $M$ denotes the size of the MGC. The main idea here is to apply the 
controlled gate, $\Lambda_1( U)=\proj{0}\otimes \one+\proj{1}\otimes U$, with $U=S^{-1}RS\trans{R}$, to the input state 
$\ket{+}\ket{0}^{\otimes \log(N)}$. Measuring the operator $X$ on the first system leads to the desired result, 
$\bra{0}RSR^{T}\ket{1}$. Due to the fact that the required classical side-computation can be performed on $\log$-space, the 
computation is performed by the exponentially smaller quantum computer.

\subsection{Ising Hamiltonian and matchgates}
	\label{sect.Preliminaries-Hamiltonian}
	
We now focus our attention on the one-dimensional Ising Hamiltonian with transverse magnetic field, and recall how the ground 
state of this Hamiltonian for various values of its parameters can be obtained via a product of matchgates acting on 
$\ket{\bf{0}}$~\cite{Kraus:11, Boyajian:13, Verstraete:09}. 

The one-dimensional Ising Hamiltonian with transverse magnetic field,
\be
\ba
H(J,B)&=-J\sum_{j=0}^{N-1}X_{j}X_{j+1} - B\sum_{j=0}^{N-1}Z_{j}\\
&\equiv -JH_1-BH_0
\ea
\label{eq. Hamiltonian}
\ee
describes a one-dimensional chain of spins with nearest-neighbour coupling interaction of strength $J$, and a global magnetic 
field $B$. Here, $X_{N}\equiv \widetilde{Z} X_0$, with $\widetilde{Z}\equiv \bigotimes _{j=0}^{N-1} Z_j$, which corresponds to 
Jordan-Wigner (JW) boundary conditions. Using the JW representation, \eq{Majorana operators}, and defining the fermionic 
operators $c_j\equiv\frac{1}{2}(x_{2j}+ix_{2j+1})$ satisfying  the fermionic commutation relations 
$\{ \adj{c} _j , \adj{c} _k \}= \{ c_j , c_k \} = 0 $ and $\{c_j, \adj{c}_k\}=\delta _{jk} $, the Hamiltonian of \eq{ Hamiltonian}, can be 
explicitly diagonalized to~\cite{Verstraete:09}
\be
\label{eq.H a}
H[a] = \sum_{j=0} ^{N-1} \epsilon_j \left( \adj{a}_j a_j -\frac{1} {2} \right),
\ee
after first performing the Fourier transform
\be
b_j=\frac{1}{\sqrt{N}}\sum_{k=0}^{N-1}\,e^{-i\frac{2\pi j k}{N}}c_k,
\label{eq.Fourier}
\ee
followed by the Bogoliubov transformation
\be
a_j=\cos \left(\frac{\theta_j}{2}\right)\, b_j -i\sin\left(\frac{\theta_j}{2}\right)\,b^\dagger_{-j}\\
\label{eq.Bogoliubov}
\ee
on the fermionic operators.  Here $-j\equiv N-j$ and
\be
\ba
\cos \left( \theta _ j \right) & = \frac {g - \cos \left(\xi _j \right) } { \sqrt { 1+ g^2 - 2 g \cos \left( \xi _j \right) } }, \\
\sin \left( \theta _j \right)& = - \frac { \sin \left( \xi _j \right ) } { \sqrt { 1+ g^2 - 2 g \cos \left( \xi _j \right) } }, \\
\ea
	\label{eq.cosine_sine_theta}
\ee
with $\xi_j = \frac{2\pi j }{N}$ and $g \equiv \frac{B}{J}$.  The energies $\epsilon_j$ are given by
\be
\epsilon_j = 2 J \sqrt{ 1+ g ^2 - 2 g \cos \left( \xi_j \right) }.
\label{eq.Epsilon j}
\ee

The ground state of the Hamiltonian in \eq{H a}, $\ket{\mathrm{G}}$, is the vacuum state $\ket{\Omega [a] }$ with respect to the 
$a$ operators~\footnote{Recall that the ground state satisfies $a_j \ket{\Omega [a] }=0 $, $\forall j$}. Clearly the ground state 
depends on the values of the coupling strength $J$ and transverse magnetic field $B$. To prepare the ground state for a 
particular value of $g$, one could either use the exact diagonalization presented above~\cite{Verstraete:09} or use an adiabatic 
evolution, starting from the ground state for $g=0$, and slowly evolving the latter into the ground state for any value of $g$. 
We shall briefly recall how the unitary evolution describing such an adiabatic evolution can indeed be given as a product of 
matchgates.

For our purposes, we will use the spin representation
\bea
\ket{\Psi}=\sum_{i_1,\ldots, i_N} \alpha_{i_1,\ldots,i_N} \ket{i_1,\ldots,i_N}
\eea
for the fermionic states
\bea \sum_{i_1,\ldots, i_N} \alpha_{i_1,\ldots,i_N} (a_i^\dagger)^{i_1}\ldots (a_N^\dagger)^{i_N}\ket{\Omega[a]}.
\eea
It will be convenient to write the vacuum state, $\ket{\Omega[a]}$, in terms of the $b$ modes as $\ket{ \mathrm{G} [b] }$. Due 
to Eq. (\ref{eq.Fourier}) only the modes $b_j,b_{-j}$ couple (for $j\neq 0,\frac{N}{2}$). Hence, we sort the modes (qubits) as $0, 
\frac {N} {2}, 1, N-1,\ldots, j,-j, \ldots \frac {N} {2}-1,\frac {N} {2}+1$. Having in mind the mapping between the fermionic states 
and the spin states and slightly misusing notation (as the vacuum state can not only be defined for some particular modes)
we consider now a single pair of these modes, $(j,-j)$. The vacuum state of modes $a$ is then given by \cite{Boyajian:15}
\be
\ket{ \Psi _j [b] } _{j , - j} \!=\!\left[\!\cos\!\left(\!\frac{\theta_j}{2}\!\right)\! \one + i \sin\!\left(\!\frac{\theta_j}{2}\!\right)\! \adj{b} _{j} 
\adj{b} _{-j} \! \right]\!
\ket{ \Omega [b] }_ {j, -j}
\label{eq.Psi j}
\ee
for $1 \leq j \leq \frac {N} {2} -1 $ and
\be
	\ket{\Psi_0 [b] } _{ 0 , \frac {N} {2} } =
	\begin{cases}
		\ket{\Omega [b] }_{0 , \frac{N}{2} }\quad \text {for $g \geq 1$},\\
		\adj{b} _ 0 \ket { \Omega [b] } _{ 0 , \frac {N} {2} } \quad \text {for $g \leq 1 $ }. \\
	\end{cases}
\ee

The spin state corresponding to the ground state is then given by
\bea
	\ket{ \mathrm{G} [b] } \equiv \ket{ \tilde{\Psi}_ 0 [b] } _{0, \frac {N} {2} } \bigotimes _{j=1} ^{\frac {N} {2} -1} \ket{ \tilde{\Psi}_j [b] } _{j , -j},
	\label{eq.Gb}
\eea
where $\ket{ \tilde{\Psi}_ j [b] }$ denotes the spin state corresponding to the two--mode fermionic state $\ket{ \Psi _j [b] }$.
The fact that the ground state is defined in two different ways, depending on the value of $g$, reflects the fact that at $g=1$ 
there is a level crossing between the ground state and the first excited state. This, however, does not prevent the adiabatic 
evolution from working, as the two states are of different parity. More precisely, the ground state has even parity for $g\geq 1$ 
and odd parity for $g<1$. However, as we will see, the adiabatic evolution is a product of matchgates which by construction 
preserves the parity. Hence, preparing the system initially in $\ketzero$, which has even parity, the adiabatic evolution will 
always result in a ground state belonging to the even parity subspace.

One way to construct the ground state for a given value of $g$ is to start with an easily preparable ground state for some value 
$g_0$ and use
the adiabatic theorem~\cite{Born:28, Kato:50, Friedrichs:55} to prepare the ground state for the desired value of $g$ by slowly 
varying this parameter form $g_0$ to $g$.  For example, the ground state for $H(B,J=0)$ is simply $\ketzero$, and it has been 
shown that an adiabatic
evolution from $\ketzero$ to the ground state $\ket{G}=\ket{\Psi(B,J)}$ corresponding to a non-zero value of $J$ can be 
described in terms of a MGC~\cite{Boyajian:13}, up to an error due to the Trotter approximation.  Furthermore, the resulting 
MGC can be compressed efficiently, and without additional error, onto a logarithmically smaller quantum 
computer~\cite{Kraus:11} as we briefly now explain.

Let us now use the notation
\bea
	\label{eq. H t}
	H ( B , J , t, T ) = - B H _0 - J \left ( \frac {t} {T} \right) H _1 ,
\eea
for the Ising Hamiltonian (see Eq.~\eqref{eq. Hamiltonian}). Here, $t$ and $T$ are some real positive coefficients. 
Note that $H ( B , J , T, T )=H(B,J)$ and $H ( B , J , 0, T )=-B H_0$ whose ground state is the state $\ketzero$. Due to the 
adiabatic theorem, evolving the state $\ketzero$ under the action of the Hamiltonian $H(B,J,t,T)$, with a parameter $t$ 
(interpreted as the time) varying from $0$ to $T$, would yield the ground state $\ket{\Psi(B,J)}$ of $H(B,J)$ as long as the 
ground state energy is non-degenerate and $T$ is large enough. For this model, a rough condition on the duration $T$ 
that guarantees that the total probability of a transition to any excited state is negligible reads $T\gg N^2 $ (see \cite{Murg:04} and references therein). The time evolution operator is given by
\bea
	\label{eq.U tilde}
	\widetilde{U} (B, J,T ) = \mathcal {T} \left[ \exp \left( \int _{ 0 } ^{ T } H (B , J , t , T ) \operatorname{d} t \right) \right],
\eea
where $\mathcal{T}$ is the time ordering operator.  It was shown that this unitary can be discretized by a Trotter 
decomposition into the unitary~\cite{Verstraete:09}
\bea
	\label{eq.U product}
	U ( B , J,T  ) = \prod _{ l = 0 } ^{ L } U_0 ( B ) \cdot U _1 ( J , l ) ,
\eea
where
\bea
	\label{eq.U0 U1}
		U _0 (B ) & =& e ^{ i B \Delta (T , L) H_0 }, \\
		U _1 (J , l) & =& e ^{ i J \frac {l} { L } \Delta (T, L ) H_1 } ,
	\eea
and $\Delta (T, L) \equiv  T / L+1 $. The unitary $U(B,J,T)$ equals $\widetilde{U} (B ,J ,T) $ up to an error which scales as $
\Order \left[ L \Delta (T, L) ^2  \right] $. Note that these unitaries can be written in the form given in \eq{U matchgate} with a 
quadratic Hamiltonian of the form given in \eq{H quadratic}. The ground state of $H(B,J)$ (with even parity) is then given by
\bea
	\label{eq.Psi U 0}
	\ket{\Psi (B , J ) } = U (B , J,T ) \ketzero
\eea
up to the error aforementioned. Another method to prepare the ground state would be to use the exact diagonalization 
presented before (see also~\cite{Verstraete:09, Boyajian:15}). As this diagonalization is achieved via matchgates, the ground 
state can be generated by applying the corresponding unitary to the initial state $\ket{\bf{0}}$.

\subsection{Metrology}
	\label{sect.Preliminaries-Metrology}
	
In this subsection we review the main tools and results of quantum metrology. Here, and throughout, we will adopt the 
frequentist (local) estimation scenario, where the parameter of interest is known to lie within a very narrow range of parameter 
values.

In local quantum metrology a sensing system, or systems,  is initialized in a known state, $\ket{\psi}$, and undergoes some 
dynamical evolution that imprints the parameter $g$ onto its state.  The sensing system is then measured. Repeating this 
procedure a large number of times allows one to obtain the requisite statistics from the measurement outcomes which are then 
used to extract an estimate $\hat{g}$ of the parameter.

For unbiased estimators, the precision in estimation of a local metrology protocol is quantified by the 
\defn{mean squared error}, $\delta g^2\equiv (g-\hat{g})^2$.  The goal is to minimize this quantity for a fixed number of 
resources of a given metrology protocol.   If the dynamical evolution is ``digitalized", i.e., consists of 
accessing a fixed dynamical evolution a given number of $N$ times, then the total resources for the protocol are the total 
number of calls $N$.  This situation corresponds to the well studied case of phase 
estimation where $g\in(0,\,2\pi]$~\cite{GLM04, Giovanetti:11}. Notice that \defn{sequential protocols}, where $N$ sequential 
calls to the evolution are made with a single sensing system, and a \defn{parallel protocol}, where the $N$ calls are made 
in parallel by employing $N$ probes each of which senses the evolution once, use the same total number of resources.    

For ``analog" dynamical evolutions, where the ``number of calls" is a continuous parameter $t$ corresponding to the time 
each sensing system is subjected to the evolution, the total resources of a given metrology protocol are $T=Nt$, 
where $N$ is the number of probes used.  This corresponds to the case of frequency estimation, where $g=\omega$ and $T$ 
can be controlled by the experimenter~\cite{Huelga:97}.  Observe now that in order 
for a sequential and parallel strategy to utilize the same amount of resources, the single sensing system has to undergo the 
dynamical evolution for a time $T=\sum_{n=1}^N\, \tau(n)$, where $\tau(n)$ is the time probe system $n$ undergoes the 
evolution in the parallel strategy.
     
In the absence of noise, which will be the assumption throughout this work, the dynamical evolution is given by the unitary 
operator $U_g$ ($U_{gt}$) where $U_x=e^{i x H}$, for the case of digital (analog) evolution respectively.  For \emph{local} 
Hamiltonians, $H=\sum_{i=1}^N h^{(i)}$, the use of entanglement allows for a quadratic improvement in precision, 
$\delta g^2=(\nu N^{2})^{-1}$ ($\delta g^2\geq (\nu (N t)^2)^{-1}=(\nu T^2)^{-1}$),
over the best known classical strategy, $\delta g^2=(\nu N)^{-1}$ ($\delta g^2\geq (\nu N t)^{-1}=(\nu T)^{-1}$) where $\nu$
denotes the number of repetitions of the experiment.  These are known as the Heisenberg and Standard quantum limits 
respectively.  Both limits can also be achieve by a sequential strategy as well.

In order to achieve the aforementioned limits, one must optimize over all possible initial states of the sensing system
as well as over all possible measurements.  This is done as follows; by the quantum~\cite{H76, BC:94} Cram\'{e}r-Rao 
inequality~\cite{Cramer:61},  the mean squared error is lower bounded by
\be
\delta g^2\geq\frac{1}{\nu \mathcal{I}[\rho(g)]},
\label{eq. CRB}
\ee 
where 
\be
\mathcal{I}[\rho(g)]=\mathrm{tr}[\rho(g) L_g],
\label{eq. QFI}
\ee
is the \defn{quantum Fisher information} of the state $\rho(g)$ and $L_g$ is the operator satisfying  
$\frac{\mathrm{d}\rho(g)}{\mathrm{d}g}=\frac{1}{2}(L_g\rho(g)+\rho(g)L_g)$ known as the
\emph{symmetric logarithmic derivative}. For a given initial state, $\rho$, the measurement maximizing the quantum Fisher 
information has the eigenprojectors of $L_g$ as its measurement operators. Thus, all that remains is to maximize over all 
possible initial states of the probe.   In the noiseless case, it can be shown that the optimal states are of the form 
$\ket{\psi}=\frac{1}{\sqrt{2}}\left(\ket{\lambda_\mathrm{min}}+\lambda_{\mathrm{max}}\right)$, where 
$\ket{\lambda_{\mathrm{max}(\mathrm{min})}}$ are the eigenstates of $H$ corresponding to the maximal (minimal) 
eigenvalue.

However, the optimal states and corresponding optimal measurement given above do not satisfy the conditions needed for 
MGC.  Indeed, even if the state is a computational basis state, the corresponding optimal measurement, given by the 
symmetric logarithmic derivative, for the case of the Ising Hamiltonian cannot be compressed. 
In order for the entire metrology protocol to be a MGC (and hence compressible) we require a suitable Hermitian operator, 
$A$ whose expectation value allows us to infer the parameter of interest with Heisenberg limited precision.  To that end, we will 
find it more convenient to compute the mean square error in our estimation
in a different way: using standard error-propagation. 

The error propagation formulae relates the variance of the operator $A$
with respect to the state $\ket{\psi(g)}$, $\Delta^2 A(g)\equiv\bmk{\psi(g)}{A^2}{\psi(g)}-\bmk{\psi(g)}{A}
{\psi(g)}^2$, to the squared error in the estimation of the parameter, $\delta g^2$ as
(see Appendix~\ref{sect.Error_propagation})
\be
\delta g^2=\frac{\Delta^2 A(g)}{\left|\partial_i \expect{A(g')}|_{g'=g}\right|^2}.
\label{eq. error_propagation}
\ee
It is this formula that will be most useful to us throughout the remainder of this work.

For the the particular case of the Ising Hamiltonian it has been demonstrated that in the absence of noise the 
best quantum mechanical strategy, indeed offers a quadratic improvement over the best classical strategy~\cite{Skotiniotis:15}.
Moreover, as the Ising Hamiltonian exhibits a phase transition, an alternative approach, based on the ground state overlap 
between two ground states near the phase transition point~\cite{Zanardi:06, Zanardi:07, Zanardi:08, Invernizzi:08, 
Gammelmark:13} has also been shown to yield super-classical scaling, even at non-zero temperature~\cite{Mehboudi:16}.   
As the ground states near the phase transition change drastically, this implies
that there exists a measurement for which one can estimate either $J$ or $B$ (assuming the other is known) with high 
precision around the phase transition point.  Note that, whilst in general one still needs to perform a complicated measurement, 
a scheme employing a much more experimentally friendly measurement that still achieves Heisenberg limited precision has 
been proposed~\cite{Invernizzi:08}.

In the next section we use the ideas~\cite{Zanardi:06, Zanardi:07, Zanardi:08, Invernizzi:08, Gammelmark:13} and construct an 
observable such that the entire protocol can be simulated on an exponentially smaller quantum computer.

\section{Compressed metrology of the Ising model}
	\label{sect. Compressed_metro}

In this section we shall show how to combine the ideas of compressed simulation for the Ising Hamiltonian, in order to perform 
a quantum metrology protocol for the precise estimation (i.e., at the Heisenberg limit) of the interaction strength $J$.  
Specifically, by starting from the ground state of the Ising Hamiltonian for $B=0$, we will adiabatically evolve the ground state 
until $B\approx J$, and then measure the expectation value of an adequate observable such that the entire circuit is a MGC.  
We will show that with just two probe qubits which are interacting with each other via the Ising interaction and $\log N$ 
additional qubits we are able to infer a precise estimate of $J$ as if we had used the Ising Hamiltonian for $N$ system probes.

A natural candidate for inferring the information about the coupling parameter $J$ would be the magnetization. However, as we
show in Appendix \ref{Magnetization}, measuring the magnetization would only lead to a suboptimal scaling in the
uncertainty.  We stress that this result is not in contradiction with the results of~\cite{Invernizzi:08} as here we are attempting to
infer the value of $J$ from the \defn{expectation} value of the magnetization.  In contrast~\cite{Invernizzi:08} measures in the
eigenbasis of the magnetization to obtain the corresponding probability distribution from which an efficient estimator is
constructed. Note that such a circuit could not be compressed as all qubits need to be measured at the end.

We will show now that the expectation value $\expect{\matB (g)}$, where here and in the following we use the notation 
$g=(B,J)$, with
\be
	\label{eq.matB}
	\matB \equiv \adj { b } _1 b _1,
\ee
estimates $J$ with optimal scaling (see also Appendix~\ref{Magnetization}).  In order to do so, let us compute the scaling of the 
derivative and the variance of $\matB$ as a function of $N$ around the phase transition. Using the scaling of these two 
functions and \eq{ error_propagation} we calculate the scaling of the uncertainty $\Delta g$.

The expectation value $\expect{\matB(g)}$ can be computed in an analogous way as in Appendix \ref{Magnetization}. We find 
that
\be
	\label{eq.expB}
	\expect{\mathcal{B}(g)} = \frac{1}{2} \left[  1 + \frac {\cos \left( \frac {2\pi} {N} \right) -g } { \sqrt{ 1 + g ^2 - 2 g \cos \left( \frac {2\pi} {N} \right) } } \right].
\ee
Computing the derivative with respect to $g$ leads to
\be
	\label{eq.expB prime}
	\expect {\matB ( g )}' = - \frac { \sin \left( \frac {2 \pi } {N} \right) ^2 } { 2 \left[ 1 + g ^2 - 2 g \cos \left( \frac {2\pi} {N} \right) \right] ^{ \frac {3} {2} } }.
\ee
Using $\sin(x)=x$ and $\cos(x)=1-\frac{x^2}{2}$ for small $x$, one can verify that 
$\expect{ \matB (g) }' |_ {g=1}  \sim \Order(N) $  for large $N$. Moreover, as $\expect{ \matB ^2}  = \expect {\matB}$ it follows 
that
\be
	\mathrm{Var} \left[ \matB (g) \right] = \frac { \sin \left( \frac {2 \pi} {N} \right) ^2 } { 4 \left[ 1 + g ^2 - 2 g \cos \left( \frac {2\pi} {N} \right)  \right] },
\ee
and making the same approximations as above for $g=1$ one obtains $\mathrm{Var} \left[ \matB (g) \right] |_ {g=1} \sim  1/4$.

From the scaling of the derivative and the variance of $\matB$ at $g=1$, we can then conclude that using 
$\expect{ \matB (g) }$ in order to estimate $g$ around the phase transition, yields an uncertainty
\be
	\left(\Delta g | _{\mathcal{B} , g=1 } \right)^2 = \frac{ \mathrm{Var} \left[\matB (g) \right]|_{g=1} } {\left[ \expect{ \matB (g)}'| _ { g=1 } \right] ^2 } \sim \Order \left( N ^{-2} \right),
\ee
meaning that an optimal scaling in the estimation of $g$ is achievable with this operator.

One might wonder whether it is also possible to compress the metrological protocol if we perform the optimal measurement
obtained from the eigenbasis of the symmetric logarithmic derivative (see \sect{Preliminaries-Metrology}).  To see that it cannot
simply observe that this measurements requires to project the final state onto
the computational basis of all $N$ qubits, which cannot be done in a compressed way.  In the next subsection we discuss how 
a compressible matchgate circuit can be constructed to measure $\expect{\matB (g)}$.

%%%%%%%%%%%%%%%%%%%%%%%%%%%%%%%%%%%%%%%%%%%%%%%%%%%%%%%%%%%%%%%%%%%%%%%%%%%%%%%%%%%%%%%%%%%%%%%%%%%%%%%%%%%%%%%%%%%%%%%%%%%%%%
\subsection{Compressed circuit to measure $\expect{ \matB ( B , J ) }$}
	\label{sect.Matchgate circuit}

We use here the results recalled in \sect{Preliminaries} to compress an $N$-qubit matchgate circuit whose output is 
$\expect{ \matB (g) }$. In order to make sure that this compressed circuit can be realized, we must ensure that the whole 
computation can be done employing only the Ising interaction (acting only on two qubits), and local operations, as we have only 
this interaction at our disposal.

As recalled in \sect{Preliminaries} (see also \cite{Kraus:11,Boyajian:13}), the ground state of the even-parity subspace of the 
Hamiltonian, is given by $\ket{ \Psi ( B , J) }  = U ( B , J ) \ketzero$, where the unitary $U(B,J)$ is given in \eq{U product} 
\footnote{Note that here and in the following we omit to write the dependency on $T$ and $L$, keeping in mind that all the 
results are an approximation which holds for large values of $T$ and $L$ (see 	Sec.~\ref{sect.Preliminaries} ).}. Hence, the 
expectation value of ${\cal B}(B,J)$ is given be
\be
	\label{eq.expB outcome}
	\expect{ \mathcal{B} (B,J) } =\bmk {\boldsymbol{0} } { \adj{U}(B,J)\, \matB \, U (B,J) } {\boldsymbol{0}}.
\ee

\begin{figure}[h]
	\centering
	\includegraphics[width=.4\textwidth]{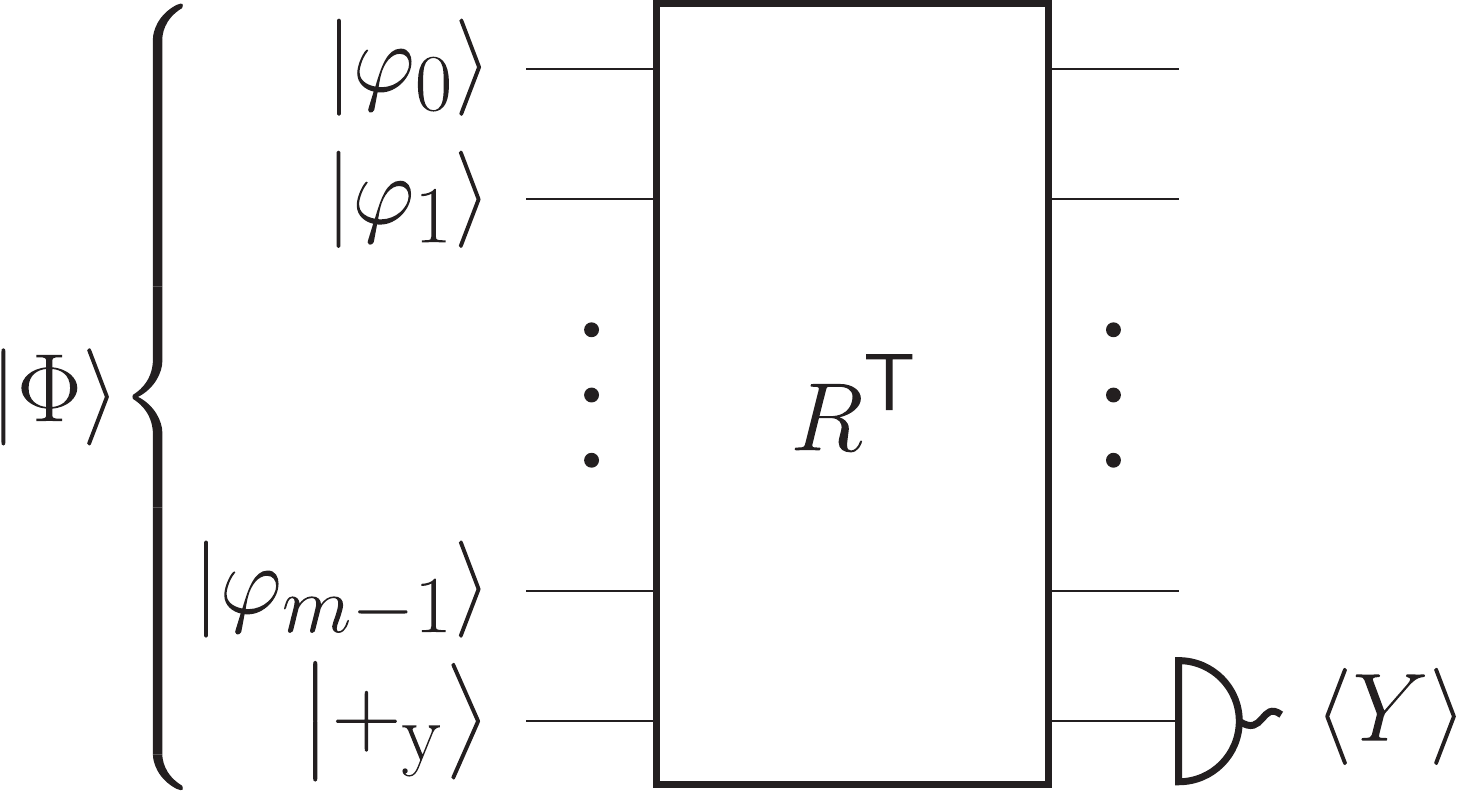}
	\caption{Schematic representation of the compressed quantum circuit which can be used to measure $\expB (B,J)$. The initial state is the $(m+1)$-qubit state $\ket{\Phi}$, which is a product state of $m$ single-qubit states $\ket{\varphi _l }=\frac{1}{\sqrt{2}} \left( \ket{0} + e^{i 2\pi 2^{l-m} }\ket{1} \right)$, for $0\leq l \leq m-1$ and one single-qubit state $\ket{+_{\mathrm{y}}}$. The initial state is transformed by an orthogonal transformation $\trans{R}$, which is decomposed into more elementary operations in \figu{R}. The circuit ends with a measurement of the last qubit on the $Y$ basis. \label{fig.Circuit}}
\end{figure}

In order to derive now a compressed circuit leading to this expectation value, we need to express $\matB$ in terms of the Majorana operators $x_j$. Using the mapping between $b$ and $c$ operators given in \eq{Fourier} and the mapping between $c$ operators and Majorana operators we find
\be
\ba
\matB [x] &= \frac {1} {4N} \sum _{ j , k = 0 } ^ { N - 1 } e ^ { i \frac { 2 \pi } { N } ( k - j ) } \left( x _{ 2 j } x _{ 2 k } + x _{ 2 j + 1 } x _{ 2 k + 1 } \right.\\
&\left. + i \, x _{ 2 j } x _{ 2 k + 1 } - i\, x _{ 2 j + 1 } x _{ 2 k } \right)\\
&= \sum _{ l , m = 0  } ^{ 2 N - 1 } b _{ l , m  } x _l x_m\,,
\ea
\label{eq.matB x}	
\ee
where
\be
	\label{eq.blm}
	b_{l,m}=
	\begin{cases}
		\frac {1} {4N} e^{i\frac{2\pi}{N}(k-j)} \quad \text {for $(l,m)=(2j,2k)$ , }\\
		\frac {i} {4N} e^{i\frac{2\pi}{N}(k-j)} \quad \text {for $(l,m)=(2j,2k+1)$ , }\\
		\frac {-i} {4N} e^{i\frac{2\pi}{N}(k-j)} \quad \text {for $(l,m)=(2j+1,2k)$ , }\\
		0 \quad \text {otherwise,}\\
	\end{cases}
\ee
for $j,k\in[0,N-1]$. Using this expression of $\matB$ it is straightforward to see that
\be
\ba
\expect{ \mathcal{B} ( B , J ) } &= \sum _{ j , k = 0 } ^ { 2 N - 1 } b _{ j , k } \bmk{ \boldsymbol { 0 } } { \adj {U} (B,J) \, x_j \,  x_k \,  U (B,J) } { \boldsymbol {0} }\\
&= -\frac{1}{2} \bmk { \Phi } { R ( B , J ) Y_m \trans{R} ( B , J ) } { \Phi },
\ea
\label{eq. expectation}
\ee
where $R(B,J)$ denotes the ``compressed gate" corresponding to $ U (B , J ) $ (see \eq{R}) and $\ket{\Phi}=\bigotimes_{l=0}^{m-1} \left( \frac{\ket{0}+e^{i2\pi 2^{l-m}}\ket{1}}{\sqrt{2}}\right)\otimes\ket{+_{\mathrm{y}}}.$ This expression represents the outcome of a quantum circuit like the one depicted in \figu{Circuit}, where the initial $(m+1)$-qubit state $\ket{\Phi}$ is transformed by the real orthogonal matrix $ \trans {R} ( B, J )$ and the operator $Y_m $ (last qubit) is measured on the output state. In \cite{Jozsa:09, Boyajian:13} the gate $R(B,J)$ has been shown to be of the form
\be
	\label{eq.R product}
	R ( B , J ) = \prod _{ l = 0 } ^ { L } R _0 ( B ) R _1 ( J , l ),
\ee
where $R _0  (B)  = e ^{ - 4 B \Delta h_0}$ and $R _1 (J,l) = e^ { - 4 J \frac {l} {L} \Delta h_1}$ correspond to $U_0 (B)$ and $U_1 (J,l)$ given in \eq{U0 U1} respectively. Here, $\Delta= T /L+1$ (see \sect{Preliminaries} ), $h_0=i\frac{1}{2}\left( \one \otimes Y_m \right)$, and $h_1=A h_0 \adj{A}$, where \be
	\label{eq.A}
	A=\sum_{j=0}^{2N-2}\kb{j+1}{j}+\kb{0}{2N-1}.
\ee

\subsection{Implementation of the compressed circuit using the Ising interaction between two qubits and local operations}

In order to implement now the compressed circuit presented above with the interactions we have at our disposal, we need to 
decompose the gates $R _0  (B)$ and $R _1 (J,l)$ into local gates and unitary evolutions corresponding to the Ising interaction. 
We will show that this can be achieved employing the Ising interaction at $B=0$ acting only on two qubits.

\begin{figure}[h]
	\centering
	\includegraphics[width=.4\textwidth]{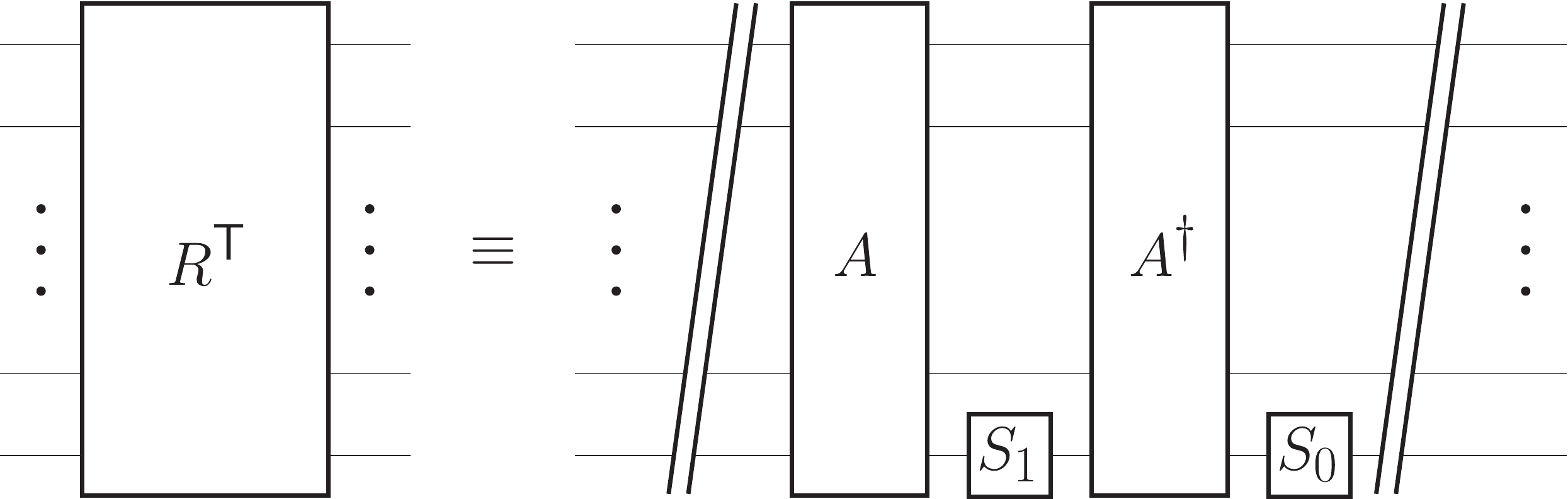}
	\caption{Here we have depicted the decomposition of one Trotter step of the matrix $\trans{R}$ into more elementary operations. The gate $\trans{R}$, whose transpose is given in \eq{R product} consists of a product of $L+1$ similar steps. The unitary operation $A$, given in \eq{A}, can be implemented as a product of $m$ controlling operations, as is shown in \figu{A}. Furthermore, the single qubit gate $S_1$, depends on the unknown parameter $J$, which we wish to estimate. In \figu{Rxx} we show how this single qubit gate can be implemented as a time evolution under the action of a two qubit Ising Hamiltonian. \label{fig.R}}
\end{figure}

Using that $ h _0  =  - i/2 \left( \one \otimes Y_m \right)$ we have $\trans{R} _0 (B) = \one \otimes S_0 (B) $, where 
$S_0 (B)= e^{ - i 2 B \Delta Y} $ is a single qubit gate acting on qubit $m$. It can be easily seen that 
$\trans{R}_1 (J,l) =  A \left( \one \otimes S_1 ( J, l)  \right) \adj{A} $, where
\be
	\label{eq.S1}
	S_1 (J,l) =e^{-i J \tau (l) Y}
\ee
with $\tau (l) = 2 l \Delta /L $~\cite{Boyajian:13}. It is important to note here that the operator $A$ does not depend on any 
parameters of the Hamiltonian. Hence, the gate $\trans{R} (B,J) $ can be decomposed into a product of terms of the form 
depicted in \figu{R}.

\begin{figure}[h]
	\centering
	\includegraphics[width=.4\textwidth]{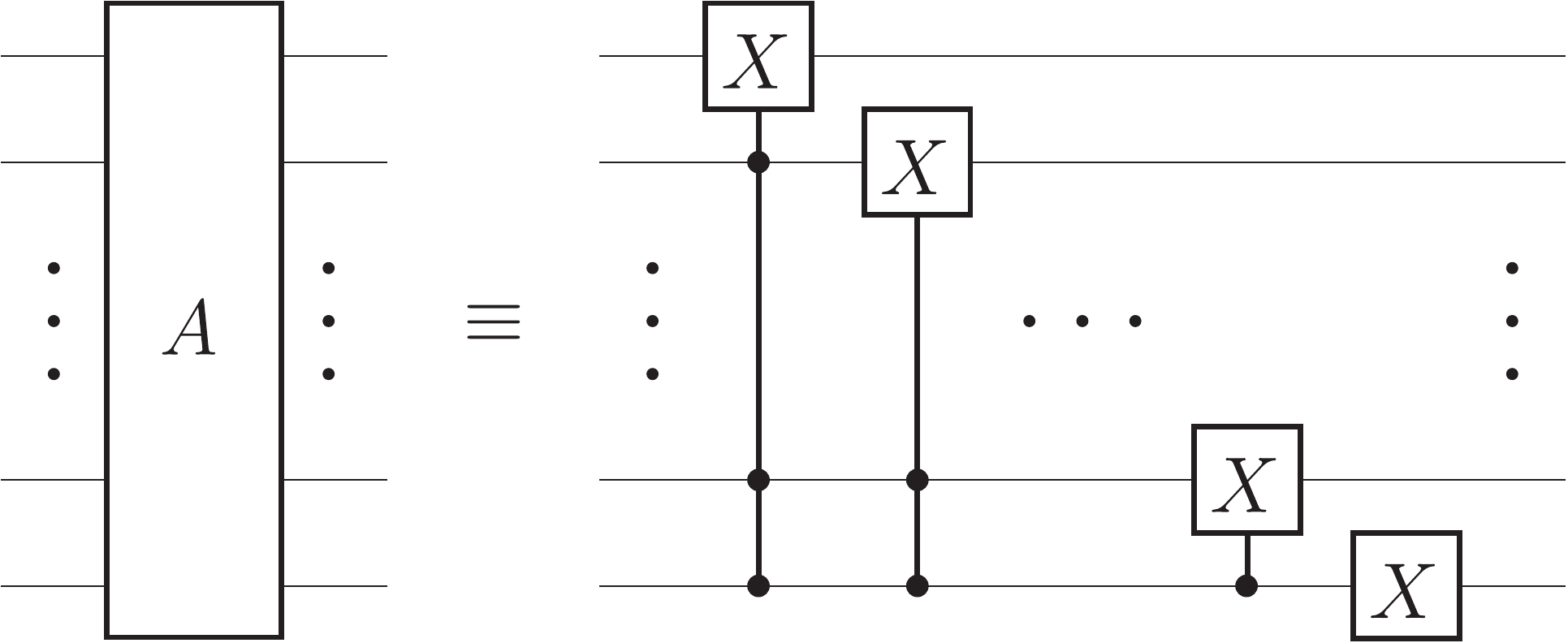}
	\caption{Decomposition of the unitary $A$ given in \eq{A decomposition} as a product of $m$ controlled-$X$ gates and a single $X$-gate acting on the last qubit. This operation corresponds to the mapping of $\ket {j}$ to $\ket{j+1\, \mathrm{mod} (N) }$ for any $j$. As $\Order (k)$ elementary gates can be used to implement a $k$-qubit controlled-$X$ operation \cite{BarBen95}, $A$ is implementable with $\Order(m^2)$ elementary gates. \label{fig.A}}
\end{figure}

Let us remark here that the operator $A$, which corresponds to the operation that maps 
$\ket {j}$ to $\ket{j+1\, \mathrm{mod} (N) }$ for any $j$, is decomposable into the simple product of controlled operations
\be
	\label{eq.A decomposition}
	A=\prod_{l=0}^{m}\Lambda_{l+1,\dots,m}(X_l),
\ee
which can then be implemented by a product of $m+1$ controlling operations, as is depicted in \figu{A}~\cite{Boyajian:13}.

The only gate which depends on the unknown parameter $J$ is the single qubit gate $S_1 (J,l) $. As we have only the Ising 
interaction (which depends of course on $J$) at our disposal, we will construct now a circuit, which simulates the action of $S_1 
(J,l)$ using $H(B,J)$ acting only on two qubits. Utilizing a single auxiliary qubit the gate $S_1 (J , l) $ can be replaced by the 
two qubit gate
\be
	R_{\mathrm{XX}}(J,l)= e^{-i J \tau (l) X\otimes X}
\ee
assisted with local operations which do not depend on $J$. As $R_{\mathrm{XX}} (J,l)$ can be implemented by letting two 
qubits evolve for a time $\tau (l)$ with interactions governed by an 2-qubit Ising Hamiltonian from the family of Hamiltonians $H 
(B,J)$ (by setting the magnetic field $B$---over which we have full control---to zero) this achieves the goal. 

\begin{figure}[h]
	\centering
	\includegraphics[width=.4\textwidth]{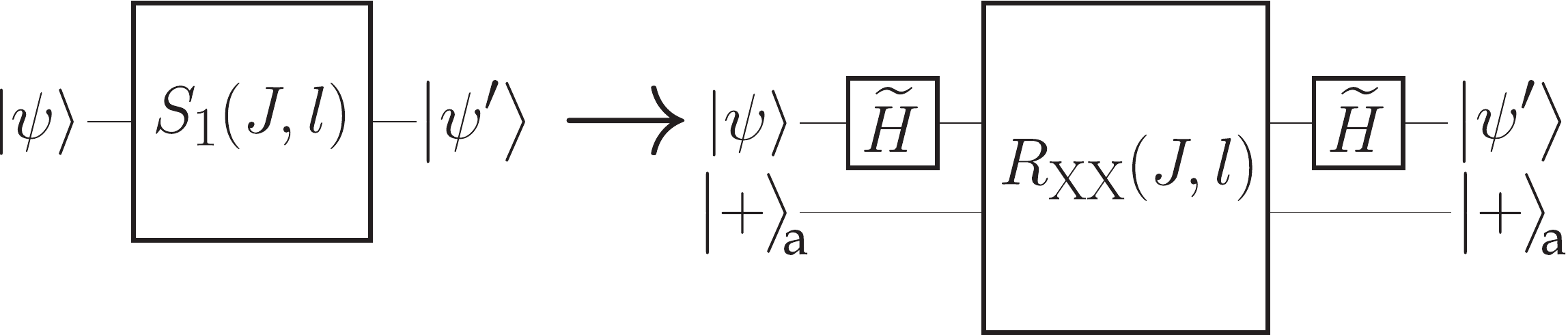}
	\caption{In this figure we depict the implementation of the single qubit $S_1 (J,l) $ using an auxiliary qubit prepared in the state $\ket{+}$ and the two-qubit gate $R_{\mathrm{XX}} (J,l)$ which can be implemented by letting two qubits evolve with interactions governed by a two-qubit Hamiltonian $H(B=0,J)$, for a time $\tau(l)= 2 l \Delta /L$.  \label{fig.Rxx} }
\end{figure}

To see this, we use an auxiliary qubit initialized in the state $\ket{+}$.  Denoting now by $\widetilde{H} \equiv \frac{1}{\sqrt{2}} \left(X+Y\right)$, the unitary for which $\widetilde{H} X\widetilde{H}=Y$ holds, we have
\be
	\ba
		S_1 \ket{j} \ket{+} _ { \mathrm {a} } & = e ^ { - i J \tau (l) Y \otimes 1 \hspace{-0.5mm} \mathrm{l}_ {\mathrm {a}} } \ket{j} \ket {+} _ { \mathrm {a} } \\
		& = \left( \widetilde{H} \otimes \one  _ { \mathrm {a} } \right) e^{-i J \tau (l) X \otimes X  _ { \mathrm {a} } } \left(\widetilde{H} \otimes \one  _ { \mathrm {a} } \right) \ket{j} \ket{+} _ { \mathrm {a} }
	\ea
\ee
for $j=0,1$, where \emph{a} denotes the auxiliary system. That is, if the auxiliary qubit is prepared in the state $\ket{+}$, then the gate $S_1 (J,l) $ can be substituted by the two-qubit gate
\be
	\label{eq.S1 tilde}
	\widetilde{S}_1 (J,l) \equiv \left( \widetilde{H} \otimes \one_ { \mathrm {a}} \right) R_{\mathrm{XX} } (J,l)  \left( \widetilde{H} \otimes \one_ { \mathrm {a}} \right),
\ee

as also depicted in \figu{Rxx}.

In summary, we have that the expectation value of ${\cal B}(B,J)$ can be measured using the following circuit.

\begin{itemize}
\item[i)]
	The system is prepared in the $(m+2)$-qubit state $\ket{\Phi}=\bigotimes_{l=0}^{m-1} \left( \frac{\ket{0}+e^{i2\pi 2^{l-m}}\ket{1}}{\sqrt{2}}\right)\otimes\ket{+_{\mathrm{y}}} \ket{+}_ {\mathrm{a}}$, where \emph{a} denotes the auxiliary system.

\item[ii)]
	The system evolves for a particular value of $J$ according to the operator $\trans{R} (B,J) $ (see \eq{R product}), where $\trans{R}_0$ is replaced by $\trans{R}_0\otimes \one_{\mathrm{a}}$ and in $R_1=A[\one \otimes S_1(J,l)]A^\dagger$, $S_1 (J,l) $ is replaced by the two-qubit gate $\widetilde{S}_1 (J,l)$, given in \eq{S1 tilde}. This is achieved by sequentially applying the gates which do not depend on the Hamiltonian parameters to the system and then letting the $m$--th qubit and the auxiliary qubit interact due to the Ising interaction (for $B=0$) and for a time $\tau (l) = 2 l \Delta /L$.

\item[iii)]
	Finally the operator $Y_m$ is measured (on qubit $m$) to retrieve the expectation value of ${\cal B}$ at the value $J$.
\end{itemize}
As shown above, this expectation value can be used to estimate the parameter $J$ optimally.

Let us now review the resources required in order to measure $\expect{ \matB (B,J) }$ using the above compressed circuit. 
The latter utilizes $m+2$ qubits upon which a sequence of gates simulating $L+1$ Trotter steps of an adiabatic evolution runs 
for a duration $T$.  Each Trotter step $l$, for $0\leq l \leq L$ can be implemented using $\Order (m^2)$ elementary gates and 
an infinitesimal time evolution step, governed by the 2-qubit Ising Hamiltonian, for a time $\tau(l)=  2l\Delta /L$, where 
$ \Delta = T/(L+1) $ .  Notice that the $\Order (m^2)$ control gates do not involve the interaction Hamiltonian in any way and 
are thus not part of the available resources. The latter are simply the total time the Ising Hamiltonian has to be used and is 
given by  $\sum_{l=0}^{L} \tau(l)=T$.  As already mentioned in \sect{ Setting} there exists a simpler strategy that also achieves 
the Heisenberg limit with the same resources $T$.  This strategy consists of preparing the two qubit state $\ket{0_x}\ket{0}$ 
and sequentially subjecting it to the modified Ising interaction 
$\left(e^{i\frac{\pi}{4}Y}\otimes \one \right) (J X_1\otimes X_2) \left(e^{i\frac{\pi}{4}Y}\otimes \one \right)$, where the control gate 
$e^{i\frac{\pi}{4}Y}$ is applied on the first system.  \

Unlike the above strategy, in order for our simulation to work we need to ensure that the digital-adiabatic evolution which we 
are using [see \eq{U product}] equals the true adiabatic evolution up to a small enough error which scales as $\Order ( L \Delta^2 ) $.  
It was shown that the choice $T\gg \Order ( N^2 ) $ guarantees that the simulated adiabatic evolution generates the ground state of the Ising 
model with probability 1~\cite{Murg:04}.  As a result the number of Trotter steps, $L$, has to be chosen such that the error in 
the discretization of the adiabatic evolution is negligible.  Moreover, this error must not exceed the error in the estimation of 
$\expect{ \matB (B,J) }$.  We find that $L=\Order(N^5)$ Trotter steps are needed to satisfy both these conditions.

We stress that even though our digital-adiabatic evolution requires $L$ calls to the Ising interaction, the duration of each call is 
such that the total time the Ising interaction is used is $T$.  This is to make sure that our simulation of the adiabatic evolution of 
the ground state followed by the measurement of $\matB$ is as faithful as possible.

\section{Conclusion}
	\label{sect.conclusion}
We have shown how compressed quantum computation can be used to simulate a quantum metrology protocol for the 
estimation of the interaction strength of a one dimensional Ising chain of $N$ spins.  Specifically, we have shown how a specific 
metrology protocol utilizing $N$ spins can be simulated on a quantum computer that uses only $\log(N)$ qubits and still 
achieves optimal precision scaling, namely $\Order(N^{-2})$.  The protocol simulated  consists of 
preparing the $N$ systems in the ground state of the Ising Hamiltonian with zero magnetic field and adiabatically evolving the 
system to the phase transition point $B\approx J$ before measuring the single fermionic mode observable $\matB$ 
of \eq{matB}.  As such a circuit is a match gate circuit, it can be efficiently compressed onto an exponentially smaller quantum 
computer.

While for the example considered here a simpler strategy yielding the same optimal precision exists, we believe that the idea of 
running a compressed metrology protocol, as outlined in this paper, can be useful when the parameter of interested has to be 
estimated from the measurement outcomes of staggered correlation functions, or for instances where the exotic states and 
measurements (and intermediate control operations) are difficult to physically implement.  

Finally, an interesting direction for future work would be the study of how different types of noise processes manifest 
themselves in the compressed protocol and whether the inclusion of additional techniques~\cite{Sekatski:16} can be utilized to 
combat noise more efficiently in the compressed protocol.  	

\paragraph*{Acknowledgements---.}This work was supported by the Austrian Science Fund (FWF): P24273-N16, P28000-N27, 
Y535-N16, SFB F40-FoQus F4012-N16 Spanish MINECO  FIS2013-40627-P, and Generalitat de Catalunya CIRIT  2014 SGR 
966.

\appendix

\section{Precision estimation using a Hermitian operator}
\label{sect.Error_propagation}

In this section we use standard error-propagation to determine the error in our estimate, $\hat{g}$, of the true parameter
$g$, when our measurement operator is equal to the Hermitian operator $A=\sum_x x\proj{x}$ and the measurement
outcome is the expectation value of $A$ with respect to the state pure state $\ket{\psi(g)}$, i.e.,
\be
	\ba
		\langle A ( g ) \rangle & = \bmk{ \psi ( g ) }  {A} { \psi ( g ) } \\
		& = \sum_x x \left| \bk { \psi ( g ) } {x} \right| ^ 2.
	\ea
	\label{eq. one}
\ee

Now let us denote the prior probability distribution over the  parameter $g$ as $p(g)$.  Then the
\defn{average value} of $\langle A(g)\rangle$
\be
\mathbb{E}_{g}\left(\langle A(g)\rangle\right)\!=\!\int\mathrm{d}_{g}p(g)\langle  A(g)\rangle
\label{eq. two}
\ee
Now consider $\mathbb{E}_{g}\left(\langle A(g')\rangle\right)$, where $g'\equiv g+\delta g\in\mathbb{R}$.  Taylor expanding the 
operator $A(g)$ to first order in $\delta g$,
\be
A(g')\!=\!A(g)+\frac{\mathrm{d} A(g')}{\mathrm{d}g'}|_{g'=g}\, \delta g,
\label{eq.three}
\ee
one obtains
\be
\mathbb{E}_{g}\left(\langle A(g')\rangle\right)\!=\!\mathbb{E}_{g}\left(\langle A(g)\rangle\right)+\frac{\mathrm{d}
A(g')}{\mathrm{d}g'}|_{g'=g}\, \epsilon,
\label{eq. four}
\ee
where $\epsilon\!\equiv\!\int p(g) \delta g\mathrm{d}g$.
It therefore follows that the \defn{average error} of the expectation value $\langle A(g)\rangle$,
\be
	\ba
		\Delta ^ 2 \langle A (g) \rangle & \equiv \int \, p ( g ) \big\{ \mathbb { E } _ { g } \left( \langle A(g) \rangle \right)- A( g+ \delta g) \big\} ^ 2  \mathrm { d } g \\
		& = \left( \frac{\mathrm{d} A (g')}{\mathrm{d}g'} | _ { g'  = g  } \right) ^ 2 \sigma ^ 2,
	\ea
	\label{eq. five}
\ee
where $\sigma^2\!\equiv\!\int p(g) \delta g^2\mathrm{d}g$ is the \defn{mean square error} of our estimation. Hence
\be
\sigma^2=\frac{\Delta^2\langle A(g)\rangle}{\left(\frac{\mathrm{d} A(g')}{\mathrm{d}g'}|_{g'=g}\right)^2}.
\label{eq. six}
\ee

Observe that for a sufficiently narrow prior, i.e., $ p(g) = \delta ( g_0-g ) $, which is the case considered in local estimation one obtains
\be
	\delta g ^ 2 = \frac{\Delta^2 A(g)} {\left(\frac{\mathrm{d} A(g')}{\mathrm{d}g'}|_{g'=g}\right)^2}
	\label{eq.seven}
\ee
where
\be
\ba
\Delta^2 A(g)&=\sum_x \left(x-\expect{ A(g) } \right)^2\, \left|\bk{x}{\psi(g)}\right|^2\\
&=\bmk{\psi(g)}{A^2}{\psi(g)}-\bmk{\psi(g)}{A}{\psi(g)}^2,
\ea
\label{eq. eight}
\ee
which is the result of \eq{ error_propagation}.

\section{Using the magnetization as an estimator of $g$ }
\label{Magnetization}

In this section we discuss the use of the magnetization of the state $\ket{\Psi}$ as an estimator of the parameter $J$ of the 
Ising Hamiltonian, i.e., the expectation value $\expect{ M (B,J) }$ with $	M=\frac{1}{N}\sum_{j=0}^{N-1}Z_j $. We compute the 
scaling of the error in the estimation of $J$ as a function of $N$.  For simplicity, in this section and in the 
following, we are going to use the variable $g=B/J$. As we assume full control of parameter $B$, estimating $g$ is 
equivalent to estimating $J$.  Furthermore, we are always going to consider measurement of observables on the state 
$\ket{\Psi}$ defined in \sect{Preliminaries-Hamiltonian}. Therefore, we are going to use the simplified notation 
$\expect{ A (B,J) } \equiv \bmk{ \Psi  } 
{A} { \Psi } $ and $\mathrm{Var ( A ) } \equiv  \expect{A^2}-\expect{A}^2$ to refer to the expectation value and the variance of an 
operator $A$ measured over the state $\ket{\Psi}$.

The motivation to choose the magnetization as an estimator of the parameter $g$ is twofold. On the one hand, a matchgate circuit to measure $\expect{ M (g) }$ can be easily constructed, and compressed in the way discussed in \sect {Preliminaries-Matchgates}. On the other hand the magnetization shows an abrupt behavior at the phase transition, which is reflected by its derivative reaching its maximum value at $g=1$. Due to the dependency of $\Delta g$ on the derivative, a large value of $\expect{M (g) }'$ allows for an estimation of $g$ with better precision about $g=1$.

In the following we compute $\expect{ M (g) }$, $\expect{ M (g) }'$ and $\mathrm{Var}\left(M (g) \right)$. Using these functions, the error $\Delta g $ can be computed according to \eq{ error_propagation}. As we will see, although the scaling of $\Delta g$ is relatively better at the phase transition, it remains sub-optimal, in the way discussed in \sect{Preliminaries-Metrology}. However, from the calculations presented in this section we are able to construct another observable which can be used instead of the magnetization operator, and from which one can estimate $g$ with optimal scaling.

Using the Jordan-Wigner transformation given in \eq{Majorana operators} and the Fourier transformation given in \eq{Fourier}, the magnetization operator $M$ can be written in term of the $b$ operators as
\be
	\label{eq.Magnetization b}
	M[b]=\frac{1}{N}\sum_{j=0}^{N-1}\left(b_j \adj{b}_j-\adj{b}_j b_j \right).
\ee
Combining this equation with the expression of the state $\ket{G[b]}$, \eq{Gb},	we have that
\be
\ba
M [b] \ket{ G [b] }&= \frac {1} {N} \left[ 2 \ket{ \tilde{\Psi} [b]} + \sum _{j=1} ^{\frac {N} {2} -1} \ket{ \Omega _0 [b]}\right.\\
&\left. \otimes  \ket{\zeta _ j [b]}  \left(  \bigotimes _{k=1,k \neq j} ^ { \frac {N} {2} -1} \ket{ \tilde{\Psi} _k[b]}  \right) \right] ,
\ea
\ee
where $\ket{\zeta _j [ b ] }  \equiv \left( b_j \adj {b } _ j - \adj{ b } _ j b _j  + b_{-j } \adj{ b } _ {-j } - \adj{ b }_ {-j } b _ {-j} \right) \ket{ \tilde{\Psi} _j [ b ] }$. It follows that the expectation value of the magnetization is given by
\be
\ba
\expect{M} &= \frac {1} {N} \left[ 2+ \sum_{j=1} ^ { \frac { N } { 2 } - 1 } \bk{ \tilde{\Psi} _j [b] } { \zeta _j [b] } \right]\\
&= \frac {2} {N} \left[ 1 + \sum_{j=1} ^{\frac {N} {2}-1} \left(1-2v_j^2\right) \right].
\ea
\ee
Recalling that $1-2 v_j^2 = \cos \left( \theta_j \right) $, given in \eq{cosine_sine_theta}, the above expression can be written explicitly as a function of $g$ as
\be
	\label{eq.Magnetization g}
	\expect { M (g) }= \frac {2} {N} \left[ 1 + \sum_{j=1} ^{ \frac { N } { 2 } -1 } \frac{ g -\cos ( \xi _j ) } { \sqrt {1+ g ^2 -2 g \cos ( \xi _j ) }} \right] .
\ee

Consequently, the derivative of the magnetization with respect to $g$ leads to the expression
\be
	\label{eq.M prime}
	\expect { M (g) }' = \frac {2} {N} \sum _{j=1} ^{ \frac {N } {2 } - 1 } \frac { \sin ^2 ( \xi _j )  } { \left[1 + g ^2 -2g \cos ( \xi _j ) \right] ^ { \frac { 3 } { 2 } } }.
\ee
Evaluating numerically for $g=1$, one can see that for large $N$ the function $\expect{M}'/\log(N)$ tends asymptotically to a constant value. Therefore we conclude that the derivative about the phase transition scales as $\expect {M (g)}'|_{g=1}  \sim \Order [ \log(N) ]$.

The variance of the magnetization, given by $\mathrm{Var} \left[ M(g) \right] = \expect{ M^2 (g) } - \expect{M(g)}^2 $ can be computed in a similar way to $\expect{M (g)}$,  to obtain
\be
\ba	
\mathrm{Var}\left[ M (g) \right] & = \frac {4} { N ^2 } \left[ 1 +\sum _{j = 1} ^{ \frac {N} {2} -1} \sin ^2 \left( \theta _j \right) \right] \\
&= \frac{4} {N ^2 } \left[ 1 + \sum _{ j = 1 } ^{ \frac {N} {2} -1 } \frac { \sin ^2 (\xi _j ) } { 1 + g^2 -2g \cos ( \xi _j ) } \right].
\ea
\label{eq.Variance M}
\ee
For large $N$ the sum in the previous expression can be approximated by an integral which for $g=1$ scales as $\Order  ( N ) $. Therefore, the variance scales as $ \mathrm{Var}\left(M (g) \right) | _ {g=1} \sim  \Order \left( N^{-1} \right)$.

In conclusion we have that using $\expect{M (g) }$ as a way to estimate the parameter $g$, yields an error in the estimation, which close to the phase transition is given by (see  \eq{ error_propagation})
\be
\ba
\left(\Delta g |_{M, g=1} \right) ^2 &= \frac { \mathrm{Var} \left[ M (g) \right] |_{g=1} }{ \left[ \expect{M (g) }' |_ {g=1} \right] ^2 } \\
& \sim \Order \left( N \log (N) \right) ^{-1}.
\ea
\ee
Hence, using the magnetization as an estimator, g can be estimated only with a suboptimal scaling on its uncertainty.

One can notice that although $\expect{ M (g)}'$ scales as $N$, some of the terms in the sum in \eq{M prime} scale more rapidly (e.g. the term with $j=1$), but they are averaged down by the global factor $2 N^{-1}$. This suggest that a different observable than $M$, where only the term with $j=1$ is considered in \eq{Magnetization b}, could be used to estimate $g$ with a better scaling than the one achieved with the magnetization, as we show in the main text.

\bibliographystyle{apsrev4-1}
\bibliography{Compressedmetro}

%merlin.mbs apsrev4-1.bst 2010-07-25 4.21a (PWD, AO, DPC) hacked
%Control: key (0)
%Control: author (72) initials jnrlst
%Control: editor formatted (1) identically to author
%Control: production of article title (-1) disabled
%Control: page (0) single
%Control: year (1) truncated
%Control: production of eprint (0) enabled
\begin{thebibliography}{38}%
\makeatletter
\providecommand \@ifxundefined [1]{%
 \@ifx{#1\undefined}
}%
\providecommand \@ifnum [1]{%
 \ifnum #1\expandafter \@firstoftwo
 \else \expandafter \@secondoftwo
 \fi
}%
\providecommand \@ifx [1]{%
 \ifx #1\expandafter \@firstoftwo
 \else \expandafter \@secondoftwo
 \fi
}%
\providecommand \natexlab [1]{#1}%
\providecommand \enquote  [1]{``#1''}%
\providecommand \bibnamefont  [1]{#1}%
\providecommand \bibfnamefont [1]{#1}%
\providecommand \citenamefont [1]{#1}%
\providecommand \href@noop [0]{\@secondoftwo}%
\providecommand \href [0]{\begingroup \@sanitize@url \@href}%
\providecommand \@href[1]{\@@startlink{#1}\@@href}%
\providecommand \@@href[1]{\endgroup#1\@@endlink}%
\providecommand \@sanitize@url [0]{\catcode `\\12\catcode `\$12\catcode
  `\&12\catcode `\#12\catcode `\^12\catcode `\_12\catcode `\%12\relax}%
\providecommand \@@startlink[1]{}%
\providecommand \@@endlink[0]{}%
\providecommand \url  [0]{\begingroup\@sanitize@url \@url }%
\providecommand \@url [1]{\endgroup\@href {#1}{\urlprefix }}%
\providecommand \urlprefix  [0]{URL }%
\providecommand \Eprint [0]{\href }%
\providecommand \doibase [0]{http://dx.doi.org/}%
\providecommand \selectlanguage [0]{\@gobble}%
\providecommand \bibinfo  [0]{\@secondoftwo}%
\providecommand \bibfield  [0]{\@secondoftwo}%
\providecommand \translation [1]{[#1]}%
\providecommand \BibitemOpen [0]{}%
\providecommand \bibitemStop [0]{}%
\providecommand \bibitemNoStop [0]{.\EOS\space}%
\providecommand \EOS [0]{\spacefactor3000\relax}%
\providecommand \BibitemShut  [1]{\csname bibitem#1\endcsname}%
\let\auto@bib@innerbib\@empty
%</preamble>
\bibitem [{\citenamefont {Huelga}\ \emph {et~al.}(1997)\citenamefont {Huelga},
  \citenamefont {Macchiavello}, \citenamefont {Pellizzari}, \citenamefont
  {Ekert}, \citenamefont {Plenio},\ and\ \citenamefont {Cirac}}]{Huelga:97}%
  \BibitemOpen
  \bibfield  {author} {\bibinfo {author} {\bibfnamefont {S.~F.}\ \bibnamefont
  {Huelga}}, \bibinfo {author} {\bibfnamefont {C.}~\bibnamefont
  {Macchiavello}}, \bibinfo {author} {\bibfnamefont {T.}~\bibnamefont
  {Pellizzari}}, \bibinfo {author} {\bibfnamefont {A.~K.}\ \bibnamefont
  {Ekert}}, \bibinfo {author} {\bibfnamefont {M.~B.}\ \bibnamefont {Plenio}}, \
  and\ \bibinfo {author} {\bibfnamefont {J.~I.}\ \bibnamefont {Cirac}},\ }\href
  {\doibase 10.1103/PhysRevLett.79.3865} {\bibfield  {journal} {\bibinfo
  {journal} {Phys. Rev. Lett.}\ }\textbf {\bibinfo {volume} {79}},\ \bibinfo
  {pages} {3865} (\bibinfo {year} {1997})}\BibitemShut {NoStop}%
\bibitem [{\citenamefont {Giovannetti}\ \emph {et~al.}(2004)\citenamefont
  {Giovannetti}, \citenamefont {Lloyd},\ and\ \citenamefont {Maccone}}]{GLM04}%
  \BibitemOpen
  \bibfield  {author} {\bibinfo {author} {\bibfnamefont {V.}~\bibnamefont
  {Giovannetti}}, \bibinfo {author} {\bibfnamefont {S.}~\bibnamefont {Lloyd}},
  \ and\ \bibinfo {author} {\bibfnamefont {L.}~\bibnamefont {Maccone}},\ }\href
  {\doibase 10.1126/science.1104149} {\bibfield  {journal} {\bibinfo  {journal}
  {Science}\ }\textbf {\bibinfo {volume} {306}},\ \bibinfo {pages} {1330}
  (\bibinfo {year} {2004})}\BibitemShut {NoStop}%
\bibitem [{\citenamefont {Giovannetti}\ \emph {et~al.}(2011)\citenamefont
  {Giovannetti}, \citenamefont {Lloyd},\ and\ \citenamefont
  {Maccone}}]{Giovanetti:11}%
  \BibitemOpen
  \bibfield  {author} {\bibinfo {author} {\bibfnamefont {V.}~\bibnamefont
  {Giovannetti}}, \bibinfo {author} {\bibfnamefont {S.}~\bibnamefont {Lloyd}},
  \ and\ \bibinfo {author} {\bibfnamefont {L.}~\bibnamefont {Maccone}},\
  }\href@noop {} {\bibfield  {journal} {\bibinfo  {journal} {Nat. Photonics}\
  }\textbf {\bibinfo {volume} {5}},\ \bibinfo {pages} {222} (\bibinfo {year}
  {2011})}\BibitemShut {NoStop}%
\bibitem [{\citenamefont {Demkowicz-Dobrza{\'n}ski}\ \emph
  {et~al.}(2015)\citenamefont {Demkowicz-Dobrza{\'n}ski}, \citenamefont
  {Jarzyna},\ and\ \citenamefont {Ko{\l}ody{\'n}ski}}]{Demkowicz:15}%
  \BibitemOpen
  \bibfield  {author} {\bibinfo {author} {\bibfnamefont {R.}~\bibnamefont
  {Demkowicz-Dobrza{\'n}ski}}, \bibinfo {author} {\bibfnamefont
  {M.}~\bibnamefont {Jarzyna}}, \ and\ \bibinfo {author} {\bibfnamefont
  {J.}~\bibnamefont {Ko{\l}ody{\'n}ski}},\ }\href@noop {} {\bibfield  {journal}
  {\bibinfo  {journal} {Prog. Optics}\ }\textbf {\bibinfo {volume} {60}},\
  \bibinfo {pages} {345} (\bibinfo {year} {2015})}\BibitemShut {NoStop}%
\bibitem [{\citenamefont {Escher}\ \emph {et~al.}(2011)\citenamefont {Escher},
  \citenamefont {de~Matos~Filho},\ and\ \citenamefont
  {Davidovich}}]{Escher:11}%
  \BibitemOpen
  \bibfield  {author} {\bibinfo {author} {\bibfnamefont {B.~M.}\ \bibnamefont
  {Escher}}, \bibinfo {author} {\bibfnamefont {R.~L.}\ \bibnamefont
  {de~Matos~Filho}}, \ and\ \bibinfo {author} {\bibfnamefont {L.}~\bibnamefont
  {Davidovich}},\ }\href {\doibase 10.1038/nphys1958} {\bibfield  {journal}
  {\bibinfo  {journal} {Nat. Phys.}\ }\textbf {\bibinfo {volume} {7}},\
  \bibinfo {pages} {406} (\bibinfo {year} {2011})}\BibitemShut {NoStop}%
\bibitem [{\citenamefont {Demkowicz-Dobrza{\'n}ski}\ \emph
  {et~al.}(2012)\citenamefont {Demkowicz-Dobrza{\'n}ski}, \citenamefont
  {Ko{\l}ody{\'n}ski},\ and\ \citenamefont {Gu{\c{t}}{\u{a}}}}]{Kolodynski:12}%
  \BibitemOpen
  \bibfield  {author} {\bibinfo {author} {\bibfnamefont {R.}~\bibnamefont
  {Demkowicz-Dobrza{\'n}ski}}, \bibinfo {author} {\bibfnamefont
  {J.}~\bibnamefont {Ko{\l}ody{\'n}ski}}, \ and\ \bibinfo {author}
  {\bibfnamefont {M.}~\bibnamefont {Gu{\c{t}}{\u{a}}}},\ }\href@noop {}
  {\bibfield  {journal} {\bibinfo  {journal} {Nat. Commun.}\ }\textbf {\bibinfo
  {volume} {3}},\ \bibinfo {pages} {1063} (\bibinfo {year} {2012})}\BibitemShut
  {NoStop}%
\bibitem [{\citenamefont {Ko{\l}ody{\'n}ski}\ and\ \citenamefont
  {Demkowicz-Dobrza{\'n}ski}(2013)}]{Kolodynski:13}%
  \BibitemOpen
  \bibfield  {author} {\bibinfo {author} {\bibfnamefont {J.}~\bibnamefont
  {Ko{\l}ody{\'n}ski}}\ and\ \bibinfo {author} {\bibfnamefont {R.}~\bibnamefont
  {Demkowicz-Dobrza{\'n}ski}},\ }\href@noop {} {\bibfield  {journal} {\bibinfo
  {journal} {New J. Phys.}\ }\textbf {\bibinfo {volume} {15}},\ \bibinfo
  {pages} {073043} (\bibinfo {year} {2013})}\BibitemShut {NoStop}%
\bibitem [{\citenamefont {Sekatski}\ \emph {et~al.}(2016)\citenamefont
  {Sekatski}, \citenamefont {Skotiniotis}, \citenamefont {Ko{\l}ody{\'n}ski},\
  and\ \citenamefont {D{\"u}r}}]{Sekatski:16}%
  \BibitemOpen
  \bibfield  {author} {\bibinfo {author} {\bibfnamefont {P.}~\bibnamefont
  {Sekatski}}, \bibinfo {author} {\bibfnamefont {M.}~\bibnamefont
  {Skotiniotis}}, \bibinfo {author} {\bibfnamefont {J.}~\bibnamefont
  {Ko{\l}ody{\'n}ski}}, \ and\ \bibinfo {author} {\bibfnamefont
  {W.}~\bibnamefont {D{\"u}r}},\ }\href@noop {} {\bibfield  {journal} {\bibinfo
   {journal} {arXiv preprint, arXiv:1603.08944}\ } (\bibinfo {year}
  {2016})}\BibitemShut {NoStop}%
\bibitem [{\citenamefont {Sachdev}(1999)}]{Sachdev:07}%
  \BibitemOpen
  \bibfield  {author} {\bibinfo {author} {\bibfnamefont {S.}~\bibnamefont
  {Sachdev}},\ }\href@noop {} {\emph {\bibinfo {title} {Quantum phase
  transitions}}}\ (\bibinfo  {publisher} {Cambridge University Press},\
  \bibinfo {year} {1999})\BibitemShut {NoStop}%
\bibitem [{\citenamefont {Kraus}(2011)}]{Kraus:11}%
  \BibitemOpen
  \bibfield  {author} {\bibinfo {author} {\bibfnamefont {B.}~\bibnamefont
  {Kraus}},\ }\href {\doibase 10.1103/PhysRevLett.107.250503} {\bibfield
  {journal} {\bibinfo  {journal} {Phys. Rev. Lett.}\ }\textbf {\bibinfo
  {volume} {107}},\ \bibinfo {pages} {250503} (\bibinfo {year}
  {2011})}\BibitemShut {NoStop}%
\bibitem [{\citenamefont {Boyajian}\ \emph {et~al.}(2013)\citenamefont
  {Boyajian}, \citenamefont {Murg},\ and\ \citenamefont {Kraus}}]{Boyajian:13}%
  \BibitemOpen
  \bibfield  {author} {\bibinfo {author} {\bibfnamefont {W.~L.}\ \bibnamefont
  {Boyajian}}, \bibinfo {author} {\bibfnamefont {V.}~\bibnamefont {Murg}}, \
  and\ \bibinfo {author} {\bibfnamefont {B.}~\bibnamefont {Kraus}},\ }\href
  {\doibase 10.1103/PhysRevA.88.052329} {\bibfield  {journal} {\bibinfo
  {journal} {Phys. Rev. A}\ }\textbf {\bibinfo {volume} {88}},\ \bibinfo
  {pages} {052329} (\bibinfo {year} {2013})}\BibitemShut {NoStop}%
\bibitem [{\citenamefont {Boyajian}\ and\ \citenamefont
  {Kraus}(2015)}]{Boyajian:15}%
  \BibitemOpen
  \bibfield  {author} {\bibinfo {author} {\bibfnamefont {W.~L.}\ \bibnamefont
  {Boyajian}}\ and\ \bibinfo {author} {\bibfnamefont {B.}~\bibnamefont
  {Kraus}},\ }\href {\doibase 10.1103/PhysRevA.92.032323} {\bibfield  {journal}
  {\bibinfo  {journal} {Phys. Rev. A}\ }\textbf {\bibinfo {volume} {92}},\
  \bibinfo {pages} {032323} (\bibinfo {year} {2015})}\BibitemShut {NoStop}%
\bibitem [{\citenamefont {Pang}\ and\ \citenamefont {Brun}(2014)}]{Pang:14}%
  \BibitemOpen
  \bibfield  {author} {\bibinfo {author} {\bibfnamefont {S.}~\bibnamefont
  {Pang}}\ and\ \bibinfo {author} {\bibfnamefont {T.~A.}\ \bibnamefont
  {Brun}},\ }\href {\doibase 10.1103/PhysRevA.90.022117} {\bibfield  {journal}
  {\bibinfo  {journal} {Phys. Rev. A}\ }\textbf {\bibinfo {volume} {90}},\
  \bibinfo {pages} {022117} (\bibinfo {year} {2014})}\BibitemShut {NoStop}%
\bibitem [{\citenamefont {Skotiniotis}\ \emph {et~al.}(2015)\citenamefont
  {Skotiniotis}, \citenamefont {Sekatski},\ and\ \citenamefont
  {D{\"u}r}}]{Skotiniotis:15}%
  \BibitemOpen
  \bibfield  {author} {\bibinfo {author} {\bibfnamefont {M.}~\bibnamefont
  {Skotiniotis}}, \bibinfo {author} {\bibfnamefont {P.}~\bibnamefont
  {Sekatski}}, \ and\ \bibinfo {author} {\bibfnamefont {W.}~\bibnamefont
  {D{\"u}r}},\ }\href@noop {} {\bibfield  {journal} {\bibinfo  {journal} {New
  J. Phys.}\ }\textbf {\bibinfo {volume} {17}},\ \bibinfo {pages} {073032}
  (\bibinfo {year} {2015})}\BibitemShut {NoStop}%
\bibitem [{\citenamefont {Zanardi}\ \emph {et~al.}(2008)\citenamefont
  {Zanardi}, \citenamefont {Paris},\ and\ \citenamefont
  {Campos~Venuti}}]{Zanardi:08}%
  \BibitemOpen
  \bibfield  {author} {\bibinfo {author} {\bibfnamefont {P.}~\bibnamefont
  {Zanardi}}, \bibinfo {author} {\bibfnamefont {M.}~\bibnamefont {Paris}}, \
  and\ \bibinfo {author} {\bibfnamefont {L.}~\bibnamefont {Campos~Venuti}},\
  }\href {\doibase 10.1103/PhysRevA.78.042105} {\bibfield  {journal} {\bibinfo
  {journal} {Phys. Rev. A}\ }\textbf {\bibinfo {volume} {78}},\ \bibinfo
  {pages} {042105} (\bibinfo {year} {2008})}\BibitemShut {NoStop}%
\bibitem [{\citenamefont {Invernizzi}\ \emph {et~al.}(2008)\citenamefont
  {Invernizzi}, \citenamefont {Korbman}, \citenamefont {Venuti},\ and\
  \citenamefont {Paris}}]{Invernizzi:08}%
  \BibitemOpen
  \bibfield  {author} {\bibinfo {author} {\bibfnamefont {C.}~\bibnamefont
  {Invernizzi}}, \bibinfo {author} {\bibfnamefont {M.}~\bibnamefont {Korbman}},
  \bibinfo {author} {\bibfnamefont {L.~C.}\ \bibnamefont {Venuti}}, \ and\
  \bibinfo {author} {\bibfnamefont {M.~G.}\ \bibnamefont {Paris}},\ }\href@noop
  {} {\bibfield  {journal} {\bibinfo  {journal} {Phys. Rev. A}\ }\textbf
  {\bibinfo {volume} {78}},\ \bibinfo {pages} {042106} (\bibinfo {year}
  {2008})}\BibitemShut {NoStop}%
\bibitem [{\citenamefont {Mehboudi}\ \emph {et~al.}(2016)\citenamefont
  {Mehboudi}, \citenamefont {Correa},\ and\ \citenamefont
  {Sanpera}}]{Mehboudi:16}%
  \BibitemOpen
  \bibfield  {author} {\bibinfo {author} {\bibfnamefont {M.}~\bibnamefont
  {Mehboudi}}, \bibinfo {author} {\bibfnamefont {L.~A.}\ \bibnamefont
  {Correa}}, \ and\ \bibinfo {author} {\bibfnamefont {A.}~\bibnamefont
  {Sanpera}},\ }\href@noop {} {\bibfield  {journal} {\bibinfo  {journal} {arXiv
  preprint, arXiv:1604.06400}\ } (\bibinfo {year} {2016})}\BibitemShut
  {NoStop}%
\bibitem [{\citenamefont {Knill}(2001)}]{Knill:01}%
  \BibitemOpen
  \bibfield  {author} {\bibinfo {author} {\bibfnamefont {E.}~\bibnamefont
  {Knill}},\ }\href@noop {} {\bibfield  {journal} {\bibinfo  {journal} {arXiv
  preprint, quant-ph/0108033}\ } (\bibinfo {year} {2001})}\BibitemShut
  {NoStop}%
\bibitem [{\citenamefont {Terhal}\ and\ \citenamefont
  {DiVincenzo}(2002)}]{Terhal:02}%
  \BibitemOpen
  \bibfield  {author} {\bibinfo {author} {\bibfnamefont {B.~M.}\ \bibnamefont
  {Terhal}}\ and\ \bibinfo {author} {\bibfnamefont {D.~P.}\ \bibnamefont
  {DiVincenzo}},\ }\href@noop {} {\bibfield  {journal} {\bibinfo  {journal}
  {Phys. Rev. A}\ }\textbf {\bibinfo {volume} {65}},\ \bibinfo {pages} {032325}
  (\bibinfo {year} {2002})}\BibitemShut {NoStop}%
\bibitem [{\citenamefont {Valiant}(2002)}]{Valiant:02}%
  \BibitemOpen
  \bibfield  {author} {\bibinfo {author} {\bibfnamefont {L.~G.}\ \bibnamefont
  {Valiant}},\ }\href@noop {} {\bibfield  {journal} {\bibinfo  {journal} {SIAM
  J. Comput.}\ }\textbf {\bibinfo {volume} {31}},\ \bibinfo {pages} {1229}
  (\bibinfo {year} {2002})}\BibitemShut {NoStop}%
\bibitem [{\citenamefont {Valiant}(2008)}]{Valiant:08}%
  \BibitemOpen
  \bibfield  {author} {\bibinfo {author} {\bibfnamefont {L.~G.}\ \bibnamefont
  {Valiant}},\ }\href@noop {} {\bibfield  {journal} {\bibinfo  {journal} {SIAM
  J. Comput.}\ }\textbf {\bibinfo {volume} {37}},\ \bibinfo {pages} {1565}
  (\bibinfo {year} {2008})}\BibitemShut {NoStop}%
\bibitem [{\citenamefont {Jozsa}\ and\ \citenamefont
  {Miyake}(2008)}]{Jozsa:08}%
  \BibitemOpen
  \bibfield  {author} {\bibinfo {author} {\bibfnamefont {R.}~\bibnamefont
  {Jozsa}}\ and\ \bibinfo {author} {\bibfnamefont {A.}~\bibnamefont {Miyake}},\
  }in\ \href@noop {} {\emph {\bibinfo {booktitle} {Proc. R. Soc. A}}},\ Vol.\
  \bibinfo {volume} {464}\ (\bibinfo {year} {2008})\ pp.\ \bibinfo {pages}
  {3089--3106}\BibitemShut {NoStop}%
\bibitem [{\citenamefont {Jozsa}\ \emph {et~al.}(2010)\citenamefont {Jozsa},
  \citenamefont {Kraus}, \citenamefont {Miyake},\ and\ \citenamefont
  {Watrous}}]{Jozsa:09}%
  \BibitemOpen
  \bibfield  {author} {\bibinfo {author} {\bibfnamefont {R.}~\bibnamefont
  {Jozsa}}, \bibinfo {author} {\bibfnamefont {B.}~\bibnamefont {Kraus}},
  \bibinfo {author} {\bibfnamefont {A.}~\bibnamefont {Miyake}}, \ and\ \bibinfo
  {author} {\bibfnamefont {J.}~\bibnamefont {Watrous}},\ }in\ \href@noop {}
  {\emph {\bibinfo {booktitle} {Proc. R. Soc. A}}},\ Vol.\ \bibinfo {volume}
  {466}\ (\bibinfo {year} {2010})\ pp.\ \bibinfo {pages} {809--830}\BibitemShut
  {NoStop}%
\bibitem [{\citenamefont {Zanardi}\ and\ \citenamefont
  {Paunkovi{\'c}}(2006)}]{Zanardi:06}%
  \BibitemOpen
  \bibfield  {author} {\bibinfo {author} {\bibfnamefont {P.}~\bibnamefont
  {Zanardi}}\ and\ \bibinfo {author} {\bibfnamefont {N.}~\bibnamefont
  {Paunkovi{\'c}}},\ }\href@noop {} {\bibfield  {journal} {\bibinfo  {journal}
  {Phys. Rev. E}\ }\textbf {\bibinfo {volume} {74}},\ \bibinfo {pages} {031123}
  (\bibinfo {year} {2006})}\BibitemShut {NoStop}%
\bibitem [{\citenamefont {Zanardi}\ \emph {et~al.}(2007)\citenamefont
  {Zanardi}, \citenamefont {Giorda},\ and\ \citenamefont
  {Cozzini}}]{Zanardi:07}%
  \BibitemOpen
  \bibfield  {author} {\bibinfo {author} {\bibfnamefont {P.}~\bibnamefont
  {Zanardi}}, \bibinfo {author} {\bibfnamefont {P.}~\bibnamefont {Giorda}}, \
  and\ \bibinfo {author} {\bibfnamefont {M.}~\bibnamefont {Cozzini}},\ }\href
  {\doibase 10.1103/PhysRevLett.99.100603} {\bibfield  {journal} {\bibinfo
  {journal} {Phys. Rev. Lett.}\ }\textbf {\bibinfo {volume} {99}},\ \bibinfo
  {pages} {100603} (\bibinfo {year} {2007})}\BibitemShut {NoStop}%
\bibitem [{Note1()}]{Note1}%
  \BibitemOpen
  \bibinfo {note} {We note that this strategy yields the exact same precision
  as the standard parallel scheme, where $N$ suitably entangled probe systems
  sense the evolution in parallel for a short time $\delta t$ before they are
  measured. Moreover the resources used are the same $N\delta t\equiv
  T$.}\BibitemShut {Stop}%
\bibitem [{\citenamefont {Verstraete}\ \emph {et~al.}(2009)\citenamefont
  {Verstraete}, \citenamefont {Cirac},\ and\ \citenamefont
  {Latorre}}]{Verstraete:09}%
  \BibitemOpen
  \bibfield  {author} {\bibinfo {author} {\bibfnamefont {F.}~\bibnamefont
  {Verstraete}}, \bibinfo {author} {\bibfnamefont {J.~I.}\ \bibnamefont
  {Cirac}}, \ and\ \bibinfo {author} {\bibfnamefont {J.~I.}\ \bibnamefont
  {Latorre}},\ }\href {\doibase 10.1103/PhysRevA.79.032316} {\bibfield
  {journal} {\bibinfo  {journal} {Phys. Rev. A}\ }\textbf {\bibinfo {volume}
  {79}},\ \bibinfo {pages} {032316} (\bibinfo {year} {2009})}\BibitemShut
  {NoStop}%
\bibitem [{Note2()}]{Note2}%
  \BibitemOpen
  \bibinfo {note} {Recall that the ground state satisfies $a_j \left | \Omega
  [a] \right >=0 $, $\forall j$}\BibitemShut {NoStop}%
\bibitem [{\citenamefont {Born}\ and\ \citenamefont {Fock}(1928)}]{Born:28}%
  \BibitemOpen
  \bibfield  {author} {\bibinfo {author} {\bibfnamefont {M.}~\bibnamefont
  {Born}}\ and\ \bibinfo {author} {\bibfnamefont {V.}~\bibnamefont {Fock}},\
  }\href@noop {} {\bibfield  {journal} {\bibinfo  {journal} {Zeitschrift
  f{\"u}r Physik}\ }\textbf {\bibinfo {volume} {51}},\ \bibinfo {pages} {165}
  (\bibinfo {year} {1928})}\BibitemShut {NoStop}%
\bibitem [{\citenamefont {Kato}(1950)}]{Kato:50}%
  \BibitemOpen
  \bibfield  {author} {\bibinfo {author} {\bibfnamefont {T.}~\bibnamefont
  {Kato}},\ }\href@noop {} {\bibfield  {journal} {\bibinfo  {journal} {Journal
  of the Physical Society of Japan}\ }\textbf {\bibinfo {volume} {5}},\
  \bibinfo {pages} {435} (\bibinfo {year} {1950})}\BibitemShut {NoStop}%
\bibitem [{\citenamefont {Friedrichs}(1955)}]{Friedrichs:55}%
  \BibitemOpen
  \bibfield  {author} {\bibinfo {author} {\bibfnamefont {K.~O.}\ \bibnamefont
  {Friedrichs}},\ }\href@noop {} {\emph {\bibinfo {title} {On the adiabatic
  theorem in quantum theory}}}\ (\bibinfo  {publisher} {Courant Institute of
  Mathematical Sciences, New York University},\ \bibinfo {year}
  {1955})\BibitemShut {NoStop}%
\bibitem [{\citenamefont {Murg}\ and\ \citenamefont {Cirac}(2004)}]{Murg:04}%
  \BibitemOpen
  \bibfield  {author} {\bibinfo {author} {\bibfnamefont {V.}~\bibnamefont
  {Murg}}\ and\ \bibinfo {author} {\bibfnamefont {J.~I.}\ \bibnamefont
  {Cirac}},\ }\href {\doibase 10.1103/PhysRevA.69.042320} {\bibfield  {journal}
  {\bibinfo  {journal} {Phys. Rev. A}\ }\textbf {\bibinfo {volume} {69}},\
  \bibinfo {pages} {042320} (\bibinfo {year} {2004})}\BibitemShut {NoStop}%
\bibitem [{\citenamefont {Helstrom}(1976)}]{H76}%
  \BibitemOpen
  \bibfield  {author} {\bibinfo {author} {\bibfnamefont {C.}~\bibnamefont
  {Helstrom}},\ }\href@noop {} {\emph {\bibinfo {title} {Quantum Detection and
  Estimation Theory}}}\ (\bibinfo  {publisher} {Academic Press, New York},\
  \bibinfo {year} {1976})\BibitemShut {NoStop}%
\bibitem [{\citenamefont {Braunstein}\ and\ \citenamefont
  {Caves}(1994)}]{BC:94}%
  \BibitemOpen
  \bibfield  {author} {\bibinfo {author} {\bibfnamefont {S.~L.}\ \bibnamefont
  {Braunstein}}\ and\ \bibinfo {author} {\bibfnamefont {C.~M.}\ \bibnamefont
  {Caves}},\ }\href@noop {} {\bibfield  {journal} {\bibinfo  {journal} {Phys.
  Rev. Lett.}\ }\textbf {\bibinfo {volume} {72}},\ \bibinfo {pages} {3439}
  (\bibinfo {year} {1994})}\BibitemShut {NoStop}%
\bibitem [{\citenamefont {Cram\'{e}r}(1961)}]{Cramer:61}%
  \BibitemOpen
  \bibfield  {author} {\bibinfo {author} {\bibfnamefont {H.}~\bibnamefont
  {Cram\'{e}r}},\ }\href@noop {} {\emph {\bibinfo {title} {Mathematical Methods
  of Statistics}}}\ (\bibinfo  {publisher} {Princeton University Press, New
  Jersey},\ \bibinfo {year} {1961})\BibitemShut {NoStop}%
\bibitem [{\citenamefont {Gammelmark}\ and\ \citenamefont
  {M{\o}lmer}(2011)}]{Gammelmark:13}%
  \BibitemOpen
  \bibfield  {author} {\bibinfo {author} {\bibfnamefont {S.}~\bibnamefont
  {Gammelmark}}\ and\ \bibinfo {author} {\bibfnamefont {K.}~\bibnamefont
  {M{\o}lmer}},\ }\href {http://stacks.iop.org/1367-2630/13/i=5/a=053035}
  {\bibfield  {journal} {\bibinfo  {journal} {New J. Phys.}\ }\textbf {\bibinfo
  {volume} {13}},\ \bibinfo {pages} {053035} (\bibinfo {year}
  {2011})}\BibitemShut {NoStop}%
\bibitem [{Note3()}]{Note3}%
  \BibitemOpen
  \bibinfo {note} {Note that here and in the following we omit to write the
  dependency on $T$ and $L$, keeping in mind that all the results are an
  approximation which holds for large values of $T$ and $L$ (see Sec.~\ref
  {sect.Preliminaries} ).}\BibitemShut {Stop}%
\bibitem [{\citenamefont {Barenco}\ \emph {et~al.}(1995)\citenamefont
  {Barenco}, \citenamefont {Bennett}, \citenamefont {Cleve}, \citenamefont
  {DiVincenzo}, \citenamefont {Margolus}, \citenamefont {Shor}, \citenamefont
  {Sleator}, \citenamefont {Smolin},\ and\ \citenamefont
  {Weinfurter}}]{BarBen95}%
  \BibitemOpen
  \bibfield  {author} {\bibinfo {author} {\bibfnamefont {A.}~\bibnamefont
  {Barenco}}, \bibinfo {author} {\bibfnamefont {C.~H.}\ \bibnamefont
  {Bennett}}, \bibinfo {author} {\bibfnamefont {R.}~\bibnamefont {Cleve}},
  \bibinfo {author} {\bibfnamefont {D.~P.}\ \bibnamefont {DiVincenzo}},
  \bibinfo {author} {\bibfnamefont {N.}~\bibnamefont {Margolus}}, \bibinfo
  {author} {\bibfnamefont {P.}~\bibnamefont {Shor}}, \bibinfo {author}
  {\bibfnamefont {T.}~\bibnamefont {Sleator}}, \bibinfo {author} {\bibfnamefont
  {J.~A.}\ \bibnamefont {Smolin}}, \ and\ \bibinfo {author} {\bibfnamefont
  {H.}~\bibnamefont {Weinfurter}},\ }\href {\doibase 10.1103/PhysRevA.52.3457}
  {\bibfield  {journal} {\bibinfo  {journal} {Phys. Rev. A}\ }\textbf {\bibinfo
  {volume} {52}},\ \bibinfo {pages} {3457} (\bibinfo {year}
  {1995})}\BibitemShut {NoStop}%
\end{thebibliography}%

 \end{document}